\pdfoutput=1
\documentclass[twocolumn,preprintnumbers,amsmath,amssymb,prx]{revtex4}
\usepackage[sort&compress]{natbib}
\usepackage[driverfallback=dvipdfm]{hyperref}
\usepackage{graphicx}
\usepackage{bm}
\usepackage{textcomp} 

\begin{document}

\title{Controlling unconventional superconductivity in artificially engineered $f$-electron Kondo superlattices
}
\author{M.\,Naritsuka\footnote{Present address: School of Physics and Astronomy, University of St Andrews, North Haugh, St Andrews, Fife KY16 9SS, UK.}}
\author{T.\,Terashima}
\author{Y.\,Matsuda}

\address{Department of Physics, Kyoto University, Kyoto 606-8502 Japan}

\begin{abstract}
Unconventional superconductivity and magnetism are intertwined on a microscopic level in a wide class of materials, including high-$T_c$ cuprates, iron pnictides, and heavy-fermion compounds.  A new approach to this most fundamental and hotly debated subject focuses on the role of interactions between superconducting electrons and bosonic fluctuations at the interface between adjacent layers in heterostructures.   
A recent state-of-the-art molecular-beam-epitaxy technique has enabled us to fabricate superlattices consisting of different heavy-fermion compounds with atomic thickness. These Kondo superlattices provide a unique opportunity to study the mutual interaction between unconventional superconductivity and magnetic order through the atomic interface.  Here, we design and fabricate hybrid Kondo superlattices consisting of alternating layers of superconducting CeCoIn$_5$ with $d$-wave pairing symmetry and nonmagnetic metal YbCoIn$_5$ or antiferromagnetic heavy fermion metals,  such as CeRhIn$_5$ and CeIn$_3$.    In these Kondo superlattices, superconducting heavy electrons are confined within the two-dimensional CeCoIn$_5$ block layers and interact with the neighboring nonmagnetic or magnetic layers through the interface.  In CeCoIn$_5$/YbCoIn$_5$ superlattices, the superconductivity is strongly influenced by the local inversion symmetry breaking at the interface.   In CeCoIn$_5$/CeRhIn$_5$ and CeCoIn$_5$/CeIn$_3$  superlattices, the superconducting and antiferromagnetic states coexist in spatially separated layers, but their mutual coupling via the interface significantly modifies the superconducting and magnetic properties.	
The fabrication of a wide variety of hybrid superlattices paves a new way to study the relationship between unconventional superconductivity and magnetism in strongly correlated materials.
\end{abstract}

\maketitle

\section{Introduction}

Superconductivity  found in several classes of strongly correlated electron systems, including cuprates\cite{Scalapino_2012,Keimer_2015}, iron pnictides/chalcogenides\cite{Hirschfeld_2011,Stewart_2017,Shibauchi_2020} and heavy fermion compounds\cite{Thalmeier_2005,Pfleiderer_2009} has attracted researchers over the past three decades.   There is almost complete consensus that  superconductivity in these systems cannot be explained by the conventional electron-phonon attractive pairing interactions \cite{Sigrist_1991,Bennemann_2008}.  As the superconductivity occurs in the vicinity of the magnetic order, it is widely believed that magnetic fluctuations, which arises from purely repulsive Coulomb interactions,  act as the source of electron pairing.  Moreover, the highest superconducting transition temperature $T_c$ is often found near a quantum critical point (QCP), at which a magnetic phase vanishes in the limit of zero temperature, indicating that  proliferation of critical magnetic excitations resulting from the QCP  plays a significant role in determining superconducting properties \cite{Mathur_1998,Park_2006,Gegenwart_2008,Knebel_2008,Nakai_2010,Hashimoto_2012,Shibauchi_2014}.  In these materials, a microscopic coexistence of superconducting and magnetically ordered phases both involving the same charge carriers is a striking example of unusual emergent electronic phases.  Despite tremendous research, however, the relationship between superconductivity and magnetism has remained largely elusive.

The strongest electron correlation is realized in heavy-fermion compounds, containing $f$ electrons (4$f$ for lanthanide and 5$f$ for actinide), especially in materials containing Ce, Pr, U and Pu atoms \cite{Coleman_2015,Stewart_1984,Joynt_2002,Huy_2007,Aoki_2019,Tsujimoto_2014,Bauer_2015}. At high temperature,  $f$ electrons are essentially localized with well-defined magnetic moments. As the temperature is lowered, the $f$ electrons begin to delocalize due to the hybridization with conduction electron band ($s$, $p$, $d$ orbital), and Kondo screening. At yet lower temperature, the $f$ electrons become itinerant, forming a narrow conduction band with heavy effective electron mass (up to a few hundred to a thousand times the free electron mass). Strong Coulomb repulsion within a narrow band and the magnetic interaction between remnant unscreened 4$f$ or 5$f$ moments leads to notable many-body effects, and superconductivity mediated by magnetic fluctuations.  The hybridized $f$ electrons are not only responsible for long-range magnetic order, but are also involved in superconductivity.  Therefore, the heavy-fermion compounds offer a fascinating playground where magnetism and unconventional superconductivity can both compete and coexist.

Among heavy fermion compounds, Ce$M$In$_5$ (where $M$ = Co, and Rh) with layered structure are ideal model systems  due to their rich electronic phase diagrams in which an intricate interplay between superconductivity and magnetism is observed \cite{D_Thompson_2012,Kenzelmann_2008,Knebel_2010}.  At ambient pressure, CeCoIn$_5$ is a superconductor ($T_c$ = 2.3\,K) with $d_{x^2-y^2}$-wave symmetry \cite{Izawa_2001,Stock_2008,An_2010,Allan_2013,Zhou_2013}. The normal state displays non-Fermi-liquid properties associated with a nearby underlying QCP \cite{Sidorov_2002,Nakajima_2007}. In contrast, CeRhIn$_5$ orders antiferromagnetically at ambient pressure ($T_N$ = 3.8\,K) \cite{Bao_2000}. Its magnetic transition is suppressed by applying pressure and the ground state becomes a purely superconducting state at $P>P\approx1.7$\,GPa, suggesting a possible presence of a pressure-induced QCP \cite{Hegger_2000,Shishido_2002,Park_2006,Knebel_2008,Park_2008,Shishido_2005,Ida_2008}.

\begin{figure}[b]
	\includegraphics[width=1.0\linewidth]{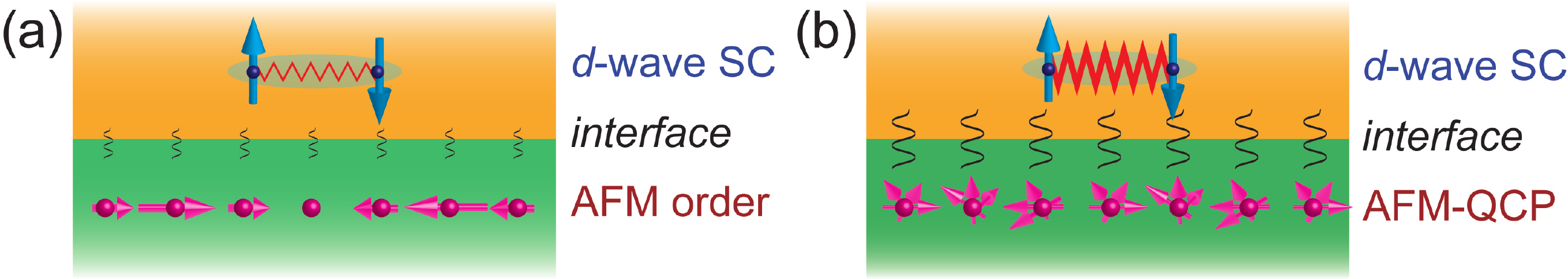}
	\caption{
		(a) Schematic figure of the interaction between $d$-wave superconducting (SC) state and static AFM order via the interface. (b) Interaction between two competing orders under pressure near a QCP, where AFM order disappears.
	}\label{Fig1}
\end{figure}

It has been shown  that interactions between superconducting electrons and bosonic excitations through an atomic interface may have a profound influence on Cooper pairing. For example, when a monolayer FeSe film grown on a SrTiO$_3$ substrate, the coupling between the FeSe electrons and SrTiO$_3$ phonons enhances the Cooper pairing, giving rise to the highest $T_c$ among all known iron-based superconductor, which is almost an order of magnitude higher than that of the bulk FeSe \cite{Shibauchi_2020,Huang_2017,Lee_2015,Lee_2014,Rademaker_2016}.  This raises the possibility of a magnetic analog in which the pairing interaction is influenced by magnetic fluctuations though an interface between an unconventional superconductor and a magnetic metal. This concept is illustrated schematically in Figs.\,\ref{Fig1}(a) and \ref{Fig1}(b). Besides allowing a new approach to revealing the entangled relationship between magnetism and unconventional superconductivity, this concept has the advantage that magnetic excitations are tunable as a magnetic transition is driven toward zero temperature, unlike phonon excitations in SrTiO$_3$.

In this review, we discuss the recent advances of Kondo superlattices which consist of alternating layers of heavy-fermion superconductor and heavy-fermion antiferromagnet.  We focus on mutual interactions between $d$-wave superconductivity  and magnetic order through the atomic interface, in particular paying  attention to how the pairing interaction is influenced by magnetic fluctuations injected  from the neighboring layers through the atomic interface.   For this purpose, we have designed and fabricated two types of  superlattices formed by alternating atomically thick layers of  CeCoIn$_5$ and  (1) conventional non-mangetic metal YbCoIn$_5$ and (2) antiferromagnetic (AFM) heavy fermion metals, such as CeRhIn$_5$ and CeIn$_3$.   In these Kondo superlattices,  superconducting heavy electrons are confined within the two-dimensional (2D)  CeCoIn$_5$ block-layers (BLs) and interact with the neighboring nonmagnetic or magnetic layers through the interface.   In CeCoIn$_5$/YbCoIn$_5$ superlattices,   local inversion symmetry breaking at the interface  enables Rashba spin-orbit coupling to play a key role in superconductivity \cite{Mizukami_2011,Goh_2012,Shimozawa_2014,Yamanaka_2015}.    In CeCoIn$_5$/CeRhIn$_5$ and CeCoIn$_5$/CeIn$_3$ Kondo superlattices,   the superconducting and AFM states coexist in spatially separated layers \cite{Naritsuka_2018,Naritsuka_2019}.  The  AFM ordering temperature of CeRhIn$_5$ and CeIn$_3$ BLs can be tuned to zero by applying hydrostatic pressure, leading to a magnetic QCP.    In these superlattices, we show  that the superconducting and non-superconducting magnetic layers can interact with each other. 
In particular, in CeCoIn$_5$/CeRhIn$_5$ superlattices, upon suppressing the AFM order, the force binding superconducting electron pairs acquires an extreme strong coupling nature  superlattices,   demonstrating that superconducting pairing can be tuned nontrivially by magnetic fluctuations injected through the interface.

\section{Kondo superlattice}
\subsection{Kondo superlattice}
Recently, the state-of-the-art molecular beam epitaxy (MBE) technique enables the realization of high quality hetero-interface of heavy fermion systems through the fabrication of Ce-based compounds \cite{Shishido_2010,Mizukami_2011,Shimozawa_2016}.  Superlattices consisting of alternating layers of superconductor CeCoIn$_5$, nonmagnetic metals, such as YbCoIn$_5$ and YbRhIn$_5$, and  AFM heavy fermion metals, such as CeRhIn$_5$ and CeIn$_3$ \cite{Naritsuka_2018,Naritsuka_2019} with atomic layer thicknesses have been fabricated.   

Here we design and fabricate three kinds of superlattices formed by alternating atomically thick layers of  CeCoIn$_5$ and (1)YbCoIn$_5$, (2)CeRhIn$_5$ and (3)CeIn$_3$.    YbCoIn$_5$ and Ce$M$In$_5$ ($M = $\,Co and Rh)  crystalize in the tetragonal HoCoGa$_5$ structure.  Ce$M$In$_5$  can be viewed as alternating layers of CeIn$_3$ and $M$In$_2$ stacked sequentially along the tetragonal $c$-axis.    CeIn$_3$ has a cubic AuCu$_3$-type structure with a 3D Fermi surface.  It should be noted that  as disorder may greatly influence physical properties, especially near a QCP, there is a great benefit in examining quantum critical systems that are stoichiometric and hence relatively disorder free; these heavy fermion compounds are examples of a small number of such systems.  

\subsection{Layered heavy-fermion CeCoIn$_5$ and CeRhIn$_5$}

CeCoIn$_5$ is a heavy-fermion superconductor with $T_c$ = 2.3 K, which is the highest among Ce-based heavy-fermion superconductors \cite{Petrovic_2001b}.  Related to the layered structure, de Haas--van Alphen experiments on CeCoIn$_5$ reveal a corrugated cylindrical Fermi surface \cite{Petrovic_2001b,Settai_2007}. In addition, nuclear magnetic resonance (NMR) relaxation rate $T_1$ measurements indicate the presence of anisotropic (quasi-2D) AFM spin fluctuations in the normal state \cite{Kawasaki_2003,Sakai_2011}.  A large Sommerfeld constant $\gamma = C/T = 290$ mJ mol$^{-1}$K$^{-2}$ is observed just above $T_c$ \cite{Ikeda_2001}. The normalized jump in heat capacity $\Delta C / \gamma T_c \sim$ 4.5 suggests that CeCoIn$_5$ exhibits very strong coupling superconductivity compared with the BCS value of 1.43 \cite{Knebel_2004}. The normal state possesses non-Fermi-liquid properties in zero fields, including $T$-linear resistivity, indicating a nearby underlying QCP  \cite{Sidorov_2002,Nakajima_2004}. It is well established by several experiments that the superconducting gap has $d_{x^2-y^2}$-wave symmetry, which is a strong indication for magnetically mediated Cooper pairing  \cite{Izawa_2001,Stock_2008,An_2010,Allan_2013,Zhou_2013}.  Inelastic neutron scattering measurements detected the presence of magnetic fluctuations at the incommensurate wavevector $q =$ (0.45, 0.45, 0.5) \cite{Raymond_2015}.   

The magnetic field destroys the superconductivity in two distinct ways, the orbital pair-breaking effect (vortex formation) and the Pauli paramagnetic effect, a breaking up of pairs by spin polarization.   In CeCoIn$_5$, the upper critical field $H_{c2}$ for both ${\bm H}\parallel ab$ and ${\bm H}\perp ab$ is limited by extremely strong Pauli pair-breaking  \cite{Izawa_2001,Bianchi_2002,Knebel_2008}.  The Pauli-limited upper critical field $H_{c2}^{\rm Pauli}$ is given by  \cite{Clogston_1962}
\begin{equation} \label{eq:Pauli_limiting_field}
	 H_{c2}^{\rm Pauli}=\sqrt{2}\Delta/g\mu_B,
\end{equation}
where $g$ is the gyromagnetic ratio, which is determined by the Ce crystalline electric field levels for CeCoIn$_5$, and $\Delta$ is the superconducting gap energy.    As a result of strong Pauli effect on the superconductivity, a possible Fulde--Ferrel--Larkin--Ovchinnikov state has been suggested \cite{Matsuda_2007,Bianchi_2003,Kumagai_2006,Kenzelmann_2008,Hatakeyama_2011,Kumagai_2011,Lin_2020}.

\begin{figure}[t]
	\centering
	\includegraphics[width=\linewidth]{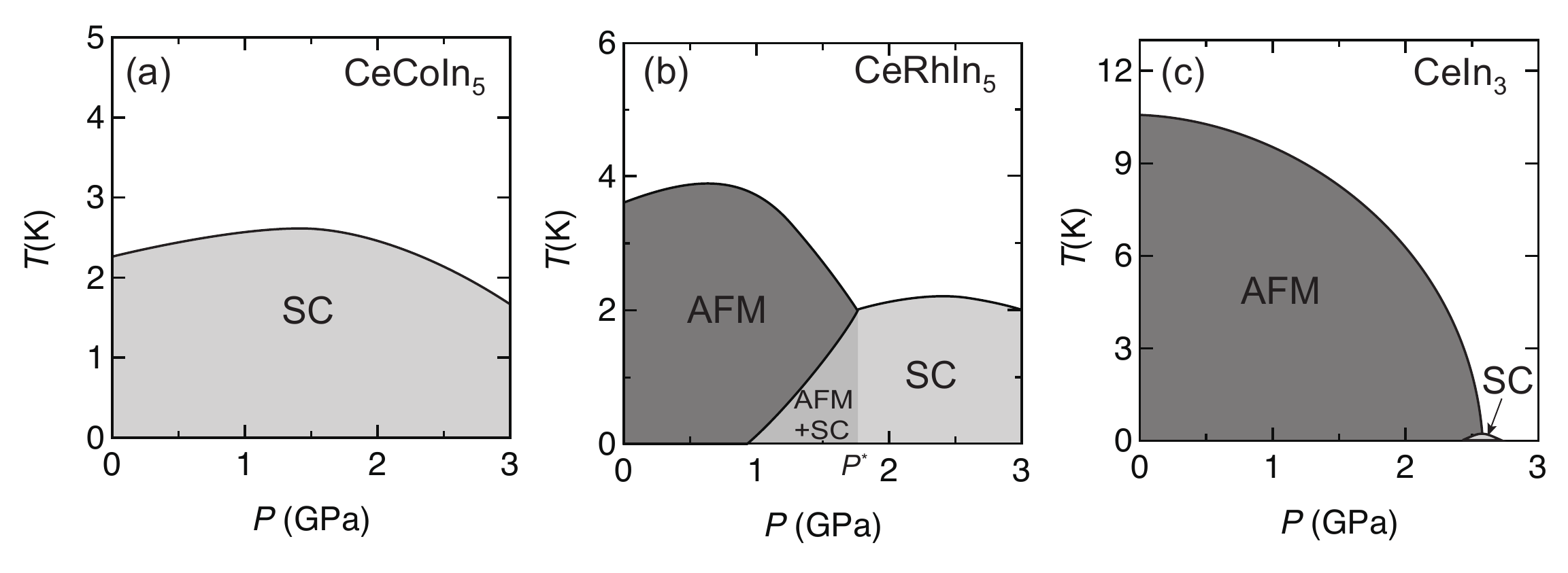}
	\caption{
		Schematic pressure-temperature phase diagrams of (a) CeCoIn$_5$, (b) CeRhIn$_5$, and (c) CeIn$_3$. Regions of antiferromagnetic order and superconducting state are indicated by AFM and SC, respectively.
	}
	\label{FigPT}
\end{figure}

At ambient pressure, superconductivity in CeCoIn$_5$ emerges from a non-Fermi liquid state. The normal state resistivity shows a linear temperature dependence , $\rho(T)\propto T$ over a wide temperature range \cite{Sidorov_2002}.  The electronic heat capacity in magnetic field, which is enough to suppress the superconductivity, exhibits a logarithmic increase , $C_e(T)/T\propto -\ln T$ \cite{Petrovic_2001b}.  These have been discussed in terms of the hallmarks of non-Fermi liquid behavior near the 2D AFM QCP \cite{L_Sarrao_2007}. With apply pressure, the Fermi liquid behavior is recovered in resistivity and heat capacity \cite{Petrovic_2001b,Sparn_2002,Knebel_2004}.  The pressure-temperature ($P$--$T$) phase diagram is displayed in Fig.\,\ref{FigPT}(a).

CeRhIn$_5$ has a smaller quasi-2D Fermi surface than that of CeCoIn$_5$ with reflecting localized nature of $f$-electrons. At ambient pressure, CeRhIn$_5$ undergoes an AFM transition at N\'{e}el temperature $T_\mathrm{N} =$ 3.8\,K, with the ordered magnetic moment of 0.75\,$\mu_\mathrm{B}$, and with an incommensurate wave vector $\bm{q}$ = (0.5, 0.5, 0.297) helical in the $c$-axis direction \cite{Bao_2000}. The $^{115}$In nuclear quadrupole resonance (NQR) measurements suggested that in the ordered state the magnetic moments lie strictly in the CeIn$_3$ plane and a spiral spin structure is realized along the $c$-axis direction \cite{Yashima_2009}.
With increasing pressure $P$, $T_\mathrm{N}$ increases and shows a maximum at $P\sim$ 1\,GPa where superconductivity appears \cite{Park_2006}. At higher pressures, the AFM phase and the superconducting phase coexist, and $T_c$ increases with pressure but $T_\mathrm{N}$ decreases \cite{Knebel_2006}. Above the pressure $P^*$ where $T_c$ = $T_\mathrm{N}$, the AFM order suddenly disappears and the pressure-induced transition to superconductivity appears to be first order, thus avoiding a QCP.  Apparent deviations from Fermi liquid behavior have been observed in $\rho(T)$ over a wide temperature and pressure range \cite{Park_2008}. The $P$--$T$ phase diagram is displayed in Fig.\,\ref{FigPT}(b).

\subsection{Antiferromagnetic heavy fermion CeIn$_3$}
At ambient pressure, CeIn$_3$  undergoes an AFM transition at $T_\mathrm{N}$ = 10.1\,K with an ordered magnetic moment of 0.48\,$\mu_\mathrm{B}$ and a commensurate wave vector $\bm{q}$ = (0.5, 0.5, 0.5) \cite{Benoit_1980,Lawrence_1980}.
A dome-like superconducting phase appears with $T_c^{\mathrm{max}}$ = 0.2\,K around the critical pressure $P_c$ = 2.6\,GPa, which indicates that the superconductivity is believed to be mediated by quantum critical spin fluctuations \cite{Mathur_1998,Lawrence_1980}.
The normal state resistivity near the critical pressure shows non-Fermi liquid behavior \cite{Knebel_2001}, $\rho(T) \sim T^\alpha$ with $\alpha = 1.6$,  which strongly deviate from the Fermi liquid value of $\alpha =$ 2. This critical exponent $\alpha$ near $P_c$ is close to 3/2 reflecting 3D AFM magnetic fluctuations \cite{Moriya_2000}, indicating the existence of 3D AFM QCP. The $P-T$ phase diagram is displayed in Fig.\,\ref{FigPT}(c).

\subsection{Non-magnetic metal YbCoIn$_5$}
The Yb-ions in YbCoIn$_5$ are divalent and form the closed-shell 4$f$ configuration \cite{Zaremba_2003}.  As a result, YbCoIn$_5$ is a  nonmagnetic compound, showing conventional metallic behavior in resistivity and magnetic susceptibility \cite{Huy_2009}. No superconducting transition has been reported in bulk and thin film YbCoIn$_5$ at ambient and under pressure \cite{Huy_2009,Naritsuka_2018}.
\section{Experimental method}

\subsection{Molecular beam epitaxy systems}
MBE is essentially a refined ultra-high-vacuum evaporation method, which helps to prevent contamination of the surface and oxidation of elements such as Ce. Thus, high-quality thin films of Ce based compounds can be grown using MBE.  MBE enables a slow growth rate of 0.01--0.02\,nm/s that permits very precise control of layer thickness. Consequently, abrupt material interfaces can be achieved, enabling the fabrication of heterostructures such as superlattices. The typical pressure in the MBE chamber is maintained at $<10^{-7}$\,Pa during the fabrication of thin films, enabling powerful diagnostic techniques such as reflection high-energy electron diffraction (RHEED) for {\it in situ} monitoring of thin films' growth without the complication of surface degradation.  

Magnesium fluoride MgF$_2$ is used as the substrate.  MgF$_2$  has a rutile-type tetragonal structure with a lattice parameter $a$ = 0.462\,nm, which matches the lattice parameters $a$ = 0.468, 0.453, 0.461 and 0.465\,nm for CeIn$_3$, YbCoIn$_5$, CeCoIn$_5$ and CeRhIn$_5$, respectively. Furthermore, because MgF$_2$ does not contain oxygen, the oxidation of Ce compounds during the growth can be avoided. Thus, single-crystal MgF$_2$ is a suitable substrate material to support the epitaxial growth of CeIn$_3$ and CeCoIn$_5$ thin films. To relax the lattice mismatch and improve the quality of the superlattice, we first grow buffer layers. Initially, 30\,nm of CeIn$_3$ buffer layers were grown at 450\textdegree{}C. Subsequently, 15\,nm of YbCoIn$_5$ buffer layers were grown at 550\textdegree{}C. On top of these, superlattice layers were grown. For CeCoIn$_5$/CeIn$_3$ superlattices, $\sim$ 5\,nm of the cobalt is deposited as a capping layer at room temperature to avoid the oxidization of samples.

In what follows, we denote a superlattice with alternating layers of $n$-UCT CeCoIn$_5$ and $m$-UCT CeRhIn$_5$ as CeCoIn$_5(n)$/CeRhIn$_5(m)$.

\begin{figure}[b]
	\centering
	\includegraphics[width=\linewidth]{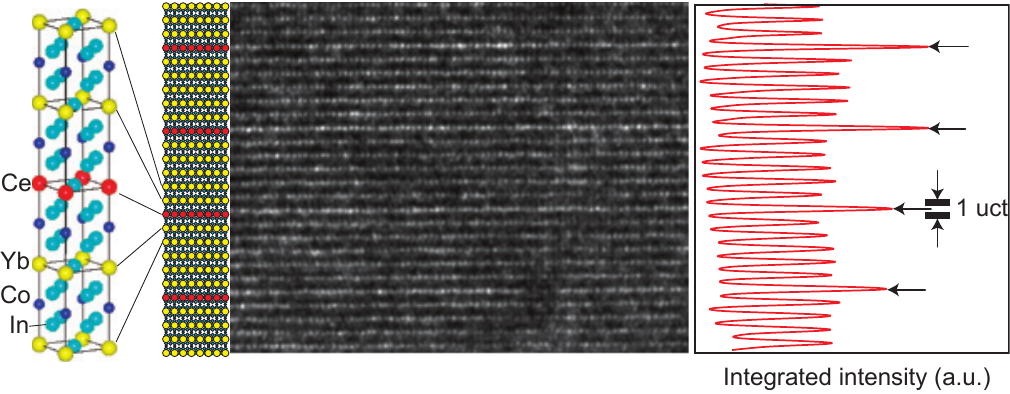}
	\caption{
		Cross-sectional TEM image of the CeCoIn$_5$(1)/YbCoIn$_5$(5) superlattice. The bright dot arrays are identified as the Ce layers and the less bright dots are Yb atoms. The right panel  represents the intensity integrated over the horizontal width of the TEM image plotted against the vertical position. 
	}
	\label{CeYb_fig1}
\end{figure}

The left panel of Fig.\,\ref{CeYb_fig1}  displays high-resolution cross-sectional  transmission electron microscope (TEM) image of CeCoIn$_5$(1)/YbCoIn$_5$(5) superlattice, where 1-UCT CeCoIn$_5$ layer are sandwiched by 5-UCT YbCoIn$_5$ layers.  The bright dot arrays are identified as the Ce layers and the less bright dots are Yb atoms, which is consistent with the designed superlattice structure.  As shown in the right panel of Fig.\,\ref{CeYb_fig1} , the intensity integrated over the horizontal width of the image plotted against the vertical position indicates a clear difference between the Ce and Yb layers, showing no discernible atomic interdiffusion between the neighboring Ce and Yb layers.

\begin{figure}[htp]
	\centering
	\includegraphics[width=0.8\linewidth]{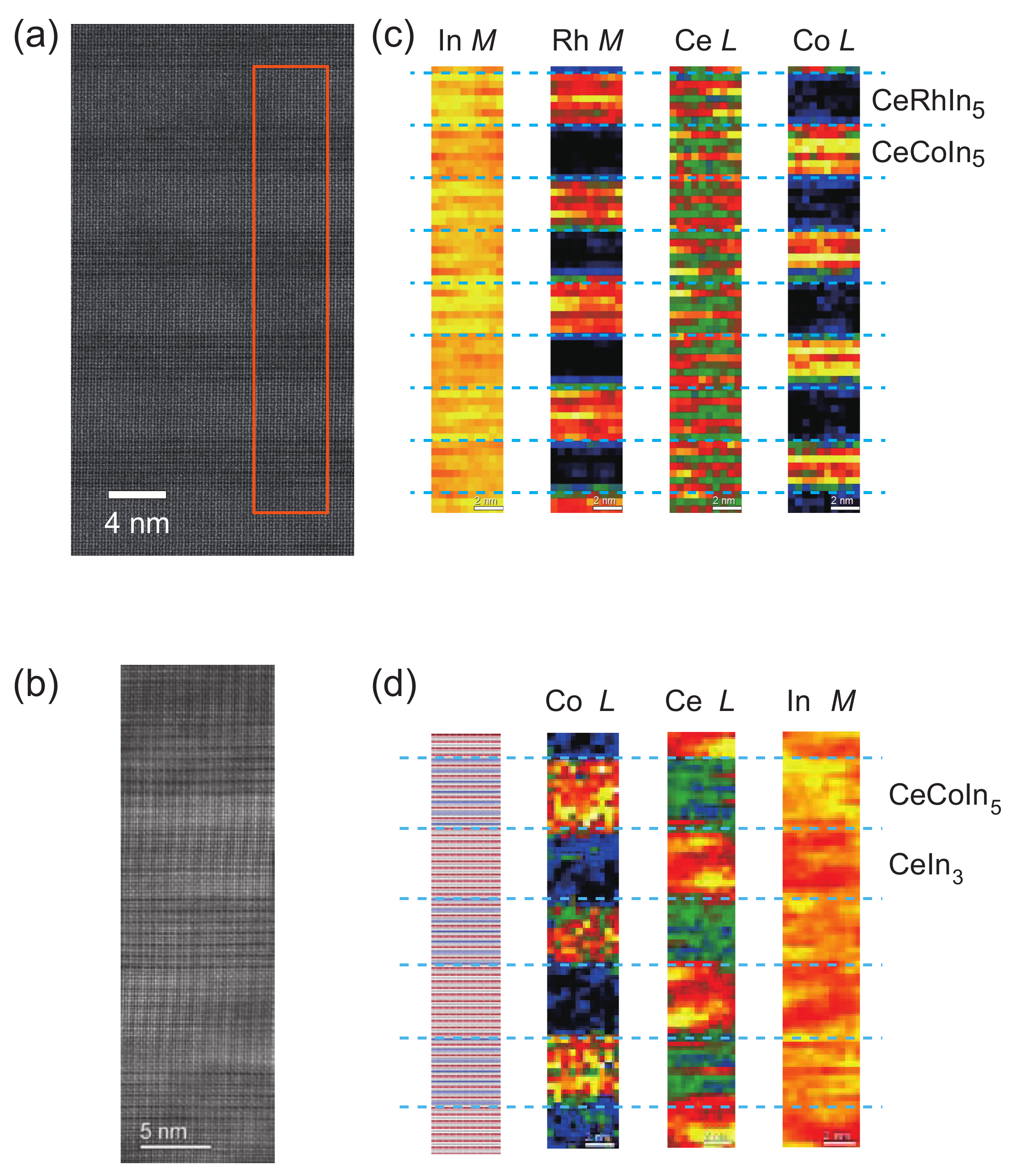}
	\caption{
		(a) and (b) represent high-resolution cross-sectional TEM and  images for CeCoIn$_5$(5)/CeRhIn$_5$(5) and CeCoIn$_5$(7)/CoIn$_3$(13)  superlattices, respectively.  (c) EELS images measured in the boxed area in the TEM image of (a) for In $M$, Rh $M$, Ce $L$, and Co $L$ edges.   (d) EELS image for the CeCoIn$_5$(7)/CoIn$_3$(13) superlattice with the electron beam alined along the (100) direction. The EELS images were taken for Co $L$, Ce $L$, and In $M$ edges. 
	}
	\label{Fig_TEM-EELS}
\end{figure}

Figures\,\ref{Fig_TEM-EELS}(a) and \ref{Fig_TEM-EELS}(b) display high resolution cross-sectional TEM images of CeCoIn$_5$(5)/CeRhIn$_5$(5) and  CeCoIn$_5$(7)/CeIn$_3$(13)  superlattices, respectively. A clear interface between CeCoIn$_5$ and CeRhIn$_5$ or CeIn$_3$ layers is observed. Figures\,\ref{Fig_TEM-EELS}(c) and \ref{Fig_TEM-EELS}(d)  display electron energy loss spectroscopy (EELS) images of  CeCoIn$_5$(5)/CeRhIn$_5$(5) and  CeCoIn$_5$(7)/CeIn$_3$(13)  superlattices, respectively. The EELS images clearly resolve CeCoIn$_5$, CeRhIn$_5$ or CeIn$_3$ BLs, demonstrating sharp interfaces with no atomic interdiffusion between the neighboring BLs.

For all the above Kondo superlattices, streak patterns of the RHEED image were observed during the whole growth of the superlattices, indicating good epitaxy.  In addition, the atomic-force-microscope measurements reveal that the surface roughness of both superlattices is within $\pm$1\,nm, which is comparable to 1--2 UCT along the $c$ axis of the constituents. Because atomically flat regions extend over distances of $\sim$0.1\,$\mu$m, it can be expected that transport properties are not seriously influenced by the roughness.    Superlattice structures were also confirmed by the satellite peaks of the x-ray diffraction patterns for all superlattices.  These results demonstrate the successful fabrication of epitaxial superlattices with sharp interfaces, revealing that the MBE technology is well suited for achieving our goal of designing Kondo superlattices.

\subsection{Pressure experiments}
In-plane resistivity measurements under pressure were performed with a commercial piston-cylinder-type high-pressure cell (CT Factory, Ltd.) \cite{Uwatoko_2003}. We used Daphne7373 oil as the pressure medium. We applied a load with a 20 tons hydraulic pressure machine at room temperature. The pressure inside the pressure cell was determined by the superconducting transition of lead (Pb), which was obtained by the quasi-four terminal resistivity measurement. A long and narrow shaped Pb installed inside the sample space is sensitive to the pressure gradient for the axial direction of the cell, which can be known by examining the width of the superconducting transition. In this study, the transition width was within 20\,mK, indicating a good hydrostatic condition.

\section{Heavy fermion superconductivity  at the metallic interface}
\subsection{2D confinement of heavy fermion superconductivity}

In CeCoIn$_5$/YbCoIn$_5$ superlattices, the Ruderman-Kittel-Kasuya-Yosida (RKKY) interaction between the Ce atoms in neighboring CeCoIn$_5$ BLs is substantially reduced \cite{Peters_2013}. Moreover, the superconducting proximity effect between CeCoIn$_5$ and YbCoIn$_5$ layers is negligibly small due to the large Fermi velocity mismatch \cite{She_2012}.  Then  an important question is whether the superconducting electrons in the superlattices are heavy and if so what their dimensionality is. When the thickness of the CeCoIn$_5$ BL is comparable to the perpendicular coherence length $\xi_c$ (about 2.1\,nm for CeCoIn$_5$), and the separation of superconducting layers ($\sim$3.7\,nm for CeCoIn$_5(n)$/YbCoIn$_5$(5) superlattices) exceeds $\xi_c$, each CeCoIn$_5$ BL acts as a 2D superconductor \cite{Mizukami_2011,Goh_2012,Schneider_1993}.   

\begin{figure}[t]
	\centering
	\includegraphics[width=\linewidth]{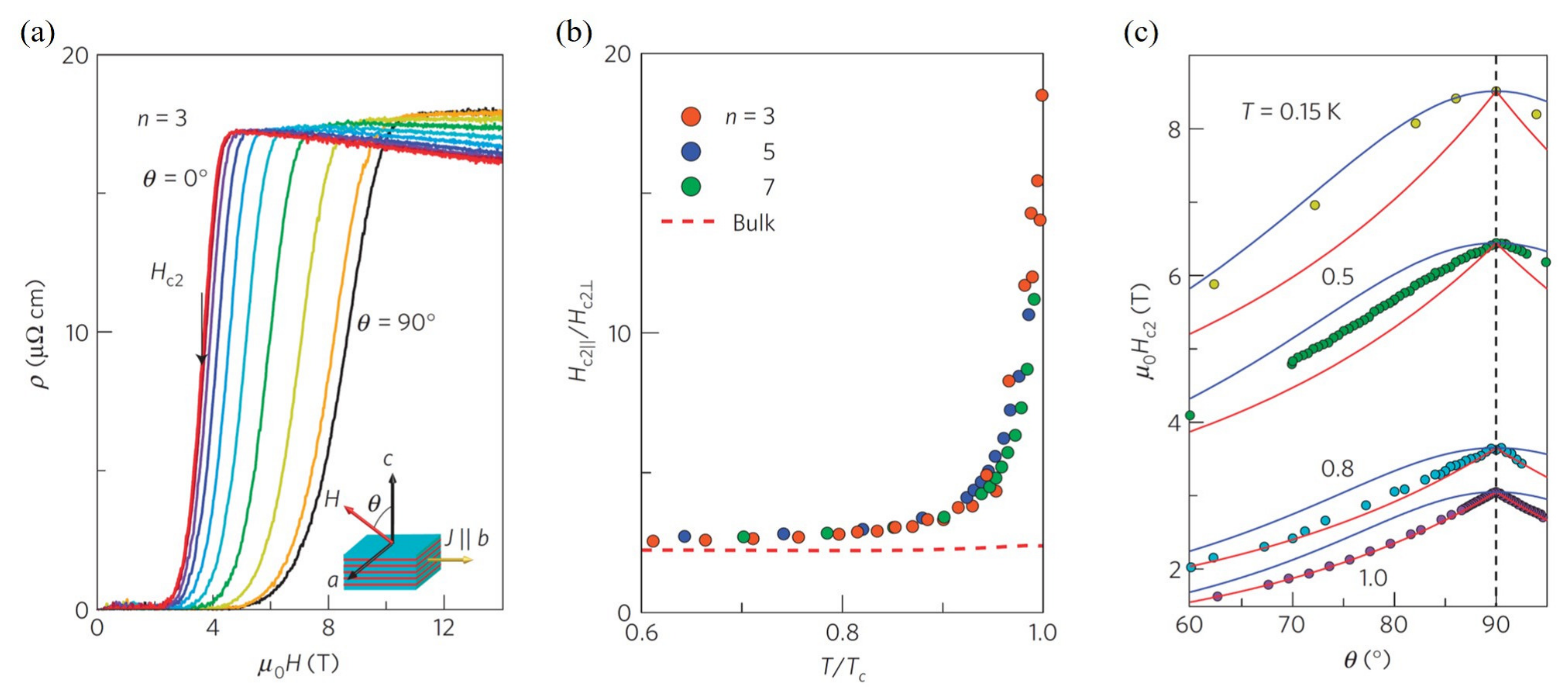}
	\caption{
	 (a) Field dependence of resistivity for CeCoIn$_5$(3)/YbCoIn$_5$(5) at several field angles from {\boldmath $H$}$\perp ab$ ($\theta=0$) to {\boldmath $H$}$\parallel a$ ($\theta=90$) at $T$=150\,mK.  (b) $H_{c2\parallel}/H_{c2\perp}$ vs. $T/T_c$, for the $n$ = 3, 5, and 7 superlattices and for the bulk CeCoIn$_5$. (c) Angular dependence of the upper critical field, $H_{c2}(\theta)$. The blue and red lines represent fits to the data using the 3D anisotropic mass model (Eq.(2)) and the 2D Tinkham model (Eq.(3)), respectively.
	}
	\label{CeYb_fig2}
\end{figure}

Figure\,\ref{CeYb_fig2}(a) depicts the magnetic-field dependence of resistivity for CeCoIn$_5$(3)/YbCoIn$_5$(5) superlattice
at several field angles from $\theta=0$ ({\boldmath $H$}$\perp ab$) to 90  ({\boldmath $H$}$\parallel a$)  at $T$=150\,mK.    Figure\,\ref{CeYb_fig2}(b) shows the anisotropy of upper critical field $H_{c2\parallel}/H_{c2\perp}$, where $H_{c2\parallel}$ and $H_{c2\perp}$  are critical field parallel and perpendicular to the $ab$ plane, as a function of reduced temperature $T/T_c$  for  CeCoIn$_5$($n$)/YbCoIn$_5$(5) superlattices with $n$=3,5 and 7 and for the bulk CeCoIn$_5$.  $H_{c2 }$ is determined by the mid-point of the resistive transition.   Unlike the almost $T$-independent anisotropy seen in single crystal of CeCoIn$_5$, anisotropy in the superlattice shows a divergent increase toward $T_c$. This diverging anisotropy is characteristic of 2D superconductivity, in which $H_{c2\parallel}$ increases as $\sqrt{T_c-T}$ due to the Pauli paramagnetic limiting, but $H_{c2\perp}$ increases as $T_c-T$ due to orbital limiting near $T_c$. 

The 2D superconductivity is also confirmed from the angle variation of $H_{c2}(\theta)$  shown in Fig.\,\ref{CeYb_fig2}(c). For 3D anisotropic mass model, $H_{c2}(\theta)$ is  represented as  \cite{Tinkham_1996}
\begin{equation}
	H_{c2}(\theta)=H_{c2\parallel}/(\sin^2\theta+\gamma_m^2\cos^2\theta)^{1/2},
	\
\end{equation}
where $\gamma_m=H_{c2\parallel}/H_{c2\perp}$.  In this model,  $H_{c2}$ varies smoothly with field orientation.  We note that when $H_{c2}$ is limited by Pauli paramagnetic effect as represented by Eq.\,(1), $H_{c2}$ also varies  smoothly as a function of  $\theta$.    On the other hand, for 2D superconductor and Josephson coupled layered superconductors,  $H_{c2}(\theta)$ is  represented by Tinkham's formula as  \cite{Tinkham_1963}
\begin{equation}
	|H_{c2}(\theta)\cos\theta/H_{c2\perp}|+|H_{c2}(\theta)\sin\theta/H_{c2\parallel}|^2=1.
\end{equation}
At $\theta = 90^{\circ}$ , $H_{c2}(\theta)$ exhibits a sharp cusp.   The solid blue and red lines in Fig.\,\ref{CeYb_fig2}(c) are the fits to Eq.\,(2) and  Eq.\,(3), respectively.    A clear cusp at $\theta = 90^{\circ}$ is observed at $T$=0.8 and 1.0\,K,  indicating the 2D superconductivity.  The cusp-like behavior of $H_{c2}(\theta)$ becomes less pronounced well below $T_c$, which is the opposite trend to the $H_{c2}(\theta)$ behavior of conventional multilayer systems.  This suggests that $H_{c2}(\theta)$ at low temperatures is dominated by the Pauli effect in any field directions.   

\begin{figure}[t]
	\centering
	\includegraphics[width=0.7\linewidth]{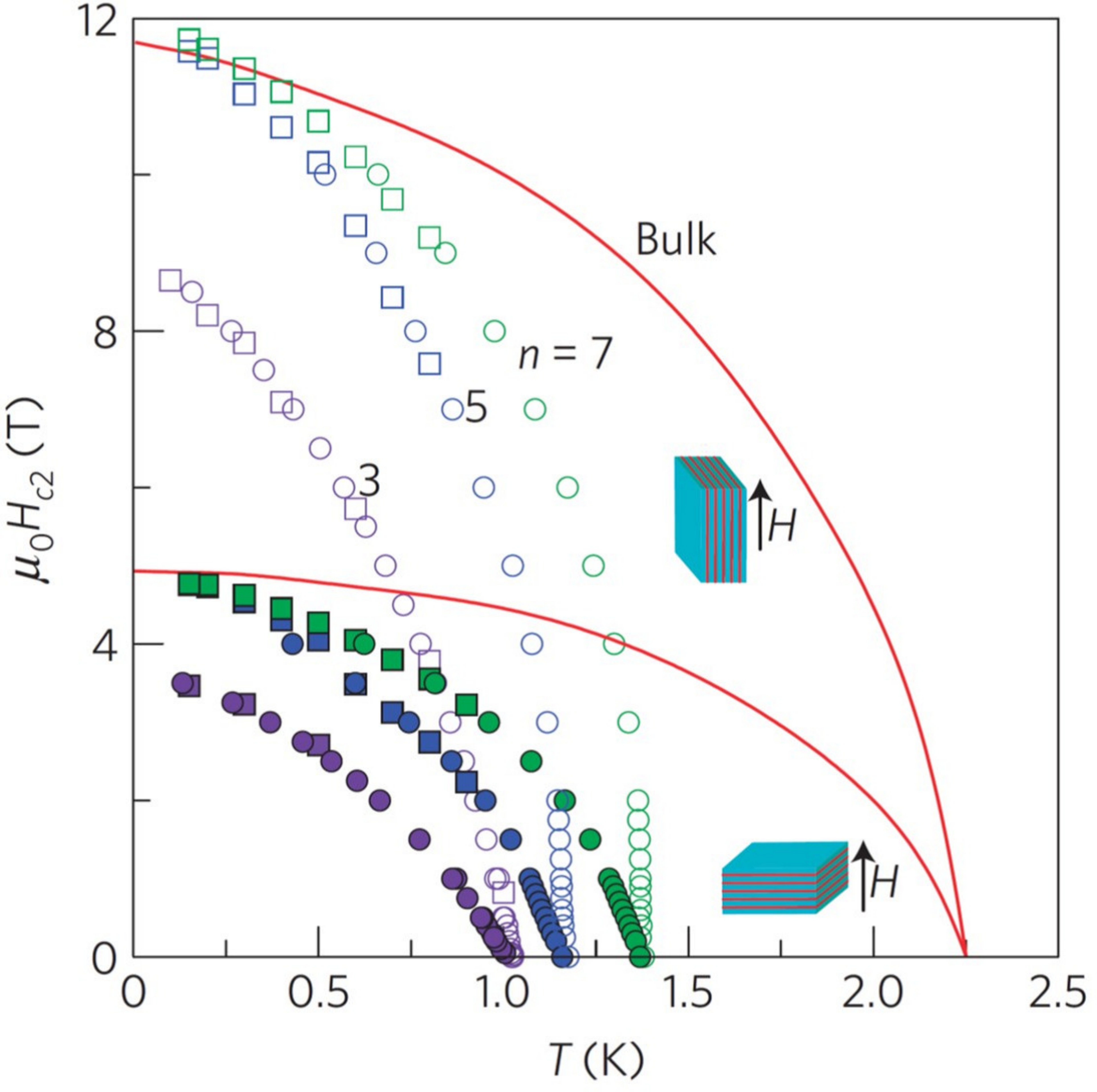}
	\caption{
		$H$--$T$ phase diagram of $n$=3, 5 and 7 superlattices in magnetic field parallel (open symbols) and perpendicular (closed symbols) to the $ab$ plane, compared to the bulk CeCoIn$_5$ data.   
	}
	\label{CeYb_fig3}
\end{figure}

Figure\,\ref{CeYb_fig3} shows $H$--$T$ phase diagram of CeCoIn$_5(n)$/YbCoIn$_5$(5) superlattices with $n$=3, 5 and 7 in magnetic field parallel (open symbols) and perpendicular (closed symbols) to the $ab$ plane.  For the comparison, the results of the bulk CeCoIn$_5$ are also plotted.   At low temperatures  $H_{c2\parallel}$ and $H_{c2\perp}$  of the superlattices  are significantly larger than those in conventional superconductors with similar $T_c$.   The zero-temperature value of the orbital upper critical field in perpendicular field $H^{\rm orb}_{c2\perp}(0)$ reflects the effective electron mass in the plane $m_{ab}^{*}$,  $H^{\rm orb}_{c2\perp}(0)\propto m_{ab}^{*2}$.   Here, $H^{\rm orb}_{c2\perp}(0)$ is determined from the initial slope of $H_{c2\perp}(T)$ at $T_c$ through the relation \cite{Werthamer_1966},
 \begin{equation}
 	H_{c2\perp}^{\rm orb}\approx 0.69T_c(-dH_{c2\perp}/dT)_{T_c}.
 \end{equation} 
$H^{\rm orb}_{c2\perp}(0)$ is estimated to be 6, 11 and 12\,T for  CeCoIn$_5(n)$/YbCoIn$_5$(5) with $n$= 3, 5 and 7, respectively. These magnitudes are
comparable with or of the same order as $H^{\rm orb}_{c2\perp}(0)$ (=14\,T) in bulk CeCoIn$_5$ \cite{Mizukami_2011}.  

Based on these results, we conclude the 2D confinement of the superconducting `heavy' electrons in the  CeCoIn$_5$/YbCoIn$_5$ superlattices.

\subsection{Local inversion symmetry breaking}

\begin{figure}[b]\centering
	\includegraphics[width=0.7\linewidth]{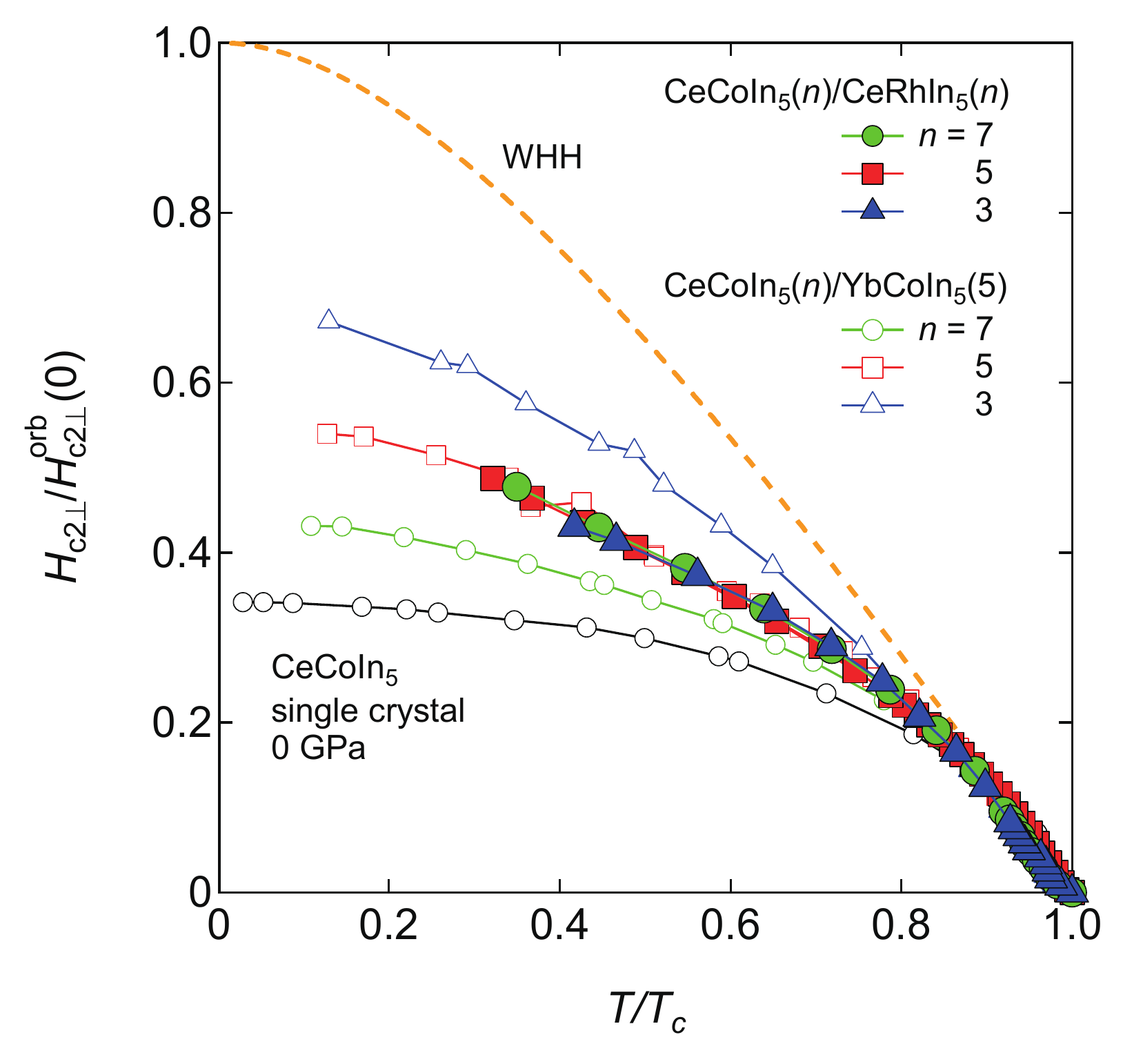}
	\caption{
		Out-of-plane upper critical field $H_{c2\perp}$ normalized by the orbital-limited upper critical field at $T$ = 0\,K, $H_{c2\perp}/H^{orb}_{c2\perp}(0)$, for   CeCoIn$_5$($n$)/YbCoIn$_5$($5$) and CeCoIn$_5$($n$)/CeRhIn$_5$($n$) superlattices with $n$ = 7, 5, and 3 are plotted as a function of the normalized temperature $T/T_c$. Two extreme cases, i.e., the result of the bulk CeCoIn$_5$ dominated by Pauli paramagnetic effect and the WHH curve with no Pauli effect, are also shown.   In CeCoIn$_5$($n$)/YbCoIn$_5$($5$) , $H_{c2\perp}/H^{orb}_{c2\perp}(0)$ is enhanced with decreasing $n$, indicating the importance of the local ISB.   In contrast,  in CeCoIn$_5$($n$)/CeRhIn$_5$($n$),  $H_{c2\perp}/H^{orb}_{c2\perp}(0)$ is independent of $n$.  
	}
	\label{Fig5}
\end{figure}

In Fig.\,\ref{Fig5}, $H_{c2\perp}$ normalized by $H_{c2\perp}^{\rm orb}(0)$ for CeCoIn$_5(n)$/YbCo$_5$(5) superlattices with $n$= 3, 5 and 7 is plotted as a function of the normalized temperature $T/T_c$.  Two extreme cases, i.e., the result of the bulk single crystal of CeCoIn$_5$ dominated by Pauli paramagnetic effect and the Werthamer-Helfand-Hohenberg (WHH) curve with no Pauli effect \cite{Werthamer_1966}, are also shown.   For all superlattices, $H_{c2\perp}/H_{c2\perp}^{\rm orb}(0)$ is much larger than that of single crystal  CeCoIn$_5$, indicating that Pauli paramagnetic pair breaking effect is reduced in these superlattices.  More importantly, $H_{c2\perp}/H_{c2\perp}^{\rm orb}(0)$  is strikingly enhanced with decreasing $n$. 

Recently, it has been suggested that the inversion symmetry breaking (ISB), together with strong spin-orbit interaction,
can dramatically affect the superconductivity\cite{Fujimoto_2007,Maruyama_2012,Bauer_2012}. It has also been pointed out that such phenomena are more pronounced in strongly correlated electron systems. The inversion symmetry imposes important constraints on the pairing states: In the presence of inversion symmetry, Cooper pairs are classified into a spin-singlet or triplet state, whereas in the absence of inversion symmetry, an asymmetric potential gradient $\nabla V$ yields a spin-orbit interaction that breaks parity, and the admixture of spin-singlet and triplet states is possible. For instance, asymmetry of the potential in the direction perpendicular to the 2D plane $\nabla V\parallel [011]$ induces Rashba spin-orbit interaction
\begin{equation}
	\alpha_R{\bm g}({\bm k})\cdot {\bm \sigma}\propto ({\bm k}\times \nabla V)\cdot {\bm \sigma},
\end{equation}
where  ${\bm g}=(-k_y,k_x,0)/k_F$, $k_F$ is the Fermi wave number, and ${\bm \sigma}$ is the Pauli matrix.
Rashba interaction splits the Fermi surface into two sheets with different spin structures \cite{Bauer_2012,Rashba_1960,Bychkov_1984}.  The energy splitting is given by $\alpha_R$, and the spin direction is tilted into the plane, rotating clockwise on one sheet and anticlockwise on the other. When the Rashba splitting exceeds the superconducting gap energy $(\alpha_R>\Delta$), the superconducting properties are dramatically modified.  As the spin-orbit interaction is generally significant in Ce-based superconductors, the introduction
of ISB makes the systems a fertile ground for observing exotic properties.   Moreover, theoretical studies suggest that when the interlayer hopping integral is comparable to, or smaller than, the Rashba splitting $(t_c \leq \alpha_R)$ , the local ISB plays an important role in determining the nature of the superconducting state\cite{Maruyama_2012}.  This appears to be the case for the CeCoIn$_5$/YbCoIn$_5$ superlattices. 

\begin{figure}[t]
	\centering
	\includegraphics[width=0.7\linewidth]{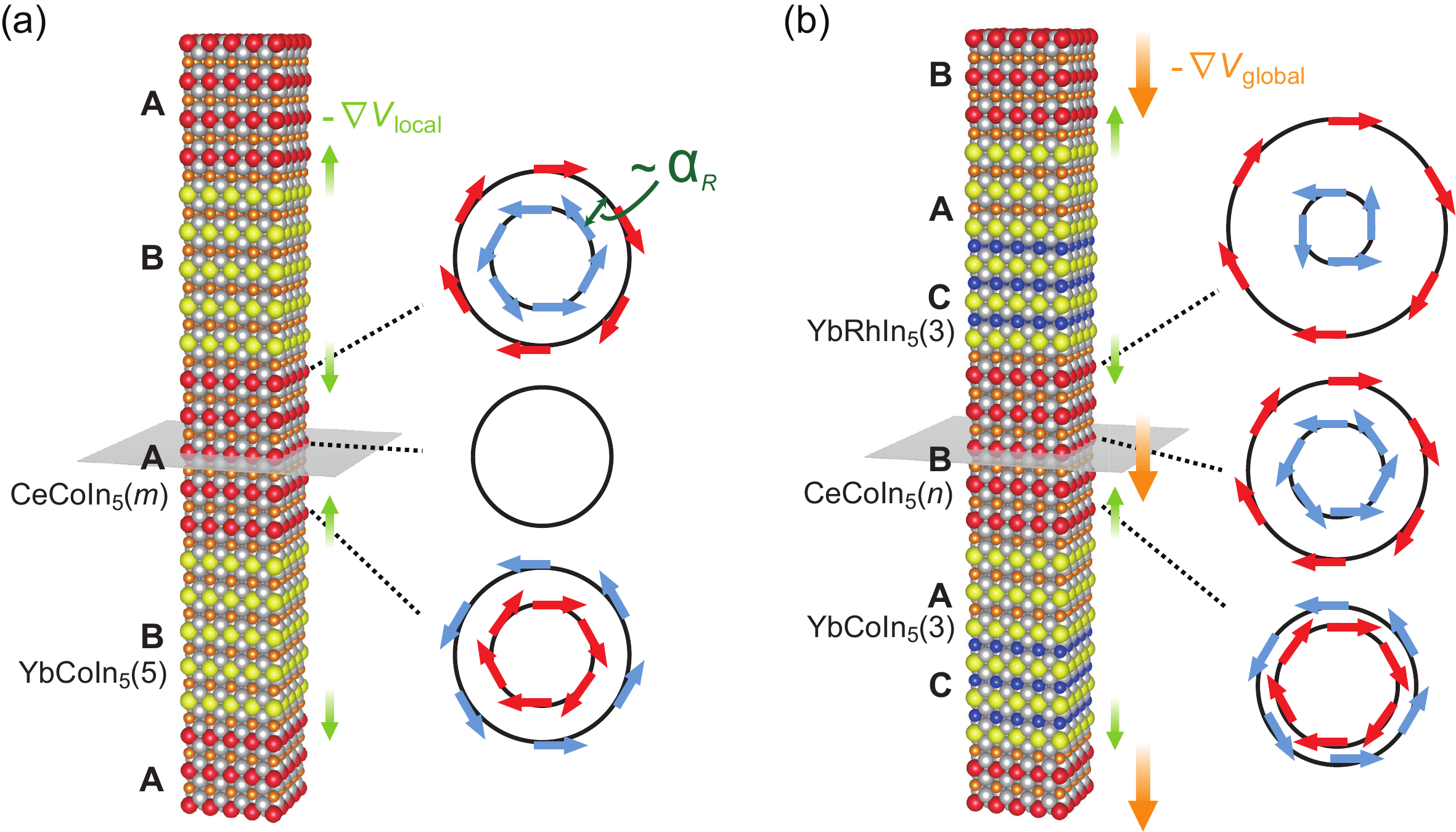}
	\caption{
		 (a) Schematic representation of bicolor Kondo superlattice CeCoIn$_5$($m$)/YbCoIn$_5$(5). The center of a CeCoIn$_5$ BL (ash plane) is a mirror plane. The green (small) arrows represent the asymmetric potential gradient associated with the local ISB, $-\nabla V_{\mathrm{local}}$. The Rashba splitting
		 occurs at the interface between the CeCoIn$_5$ and YbCoIn$_5$ BLs due to the local ISB. The spin direction is rotated in the $ab$ plane and is
		 opposite between the top and bottom CeCoIn$_5$ BLs.   (b) Schematic representation of noncentrosymmetric tricolor Kondo superlattices YbCoIn$_5$(3)/CeCoIn$_5(n)$/YbRhIn$_5$(3). The orange (large) arrows represent the asymmetric potential gradient $-\nabla V_{\mathrm{global}}$.  In the tricolor superlattices, all layers are not the mirror planes. The orange arrows represent the asymmetric potential gradient $-\nabla V_{\rm global}$ due to the global broken inversion symmetry.   The amplitude of Rashba splitting at the top layer of CeCoIn$_5$ BL is larger than that at the bottom of the BL, owing to the presence of $-\nabla V_{\mathrm{global}}$ shown by the green small arrows.  
	}
	\label{tricolor_schematic}
\end{figure}

Although these superlattices maintain centrosymmetry, it has been suggested that the local ISB at the interface between two compounds influences the superconducting.   Figure\,\ref{tricolor_schematic}(a) represents the schematic representation of CeCoIn$_5(m)$/YbCoIn$_5(5)$ superlattice.  The middle CeCoIn$_5$ layer in a given CeCoIn$_5$ BL indicated by the gray plane is a mirror plane. The green (small) arrows represent the asymmetric potential gradient associated with the local ISB,  $-\nabla V_{\mathrm{local}}$. The Rashba splitting occurs at the interface between the CeCoIn$_5$ and YbCoIn$_5$ due to the local ISB. The spin direction is rotated in the $ab$ plane and is opposite between the top and bottom CeCoIn$_5$  layers.   Because  the fraction of noncentrosymmetric interface layers increases with decreasing $n$, the observed remarkable enhancement of  $H_{c2\perp}/H_{c2\perp}^{\rm orb}(0)$ with decreasing $n$ shown in Fig.\,\ref{Fig5}  is attributed to the increased importance of the local ISB.  

It has been reported that the local ISB also seriously influence the magnetic properties in non-superconducting Kondo superlattices.  In CeRhIn$_5$/YbRhIn$_5$ superlattices, with reducing the thickness of  magnetic CeRhIn$_5$ BLs, the N\'{e}el temperature is suppressed  and the quasiparticle mass is strongly enhanced, implying dimensional control toward a magnetic QCP  \cite{Ishii_2016}. 

\subsection{Tricolor Kondo superlattices}
Recently,  it has been reported that the magnitude of the Rashba spin-orbit interaction arising from the ISB is controllable by fabricating two types of  Kondo superlattices\cite{Shimozawa_2014, Naritsuka_2017}.  One is the introduction of the thickness modulation of YbCoIn$_5$ BLs that  breaks the inversion symmetry centered at the superconducting block of CeCoIn$_5$.   The other is the `tricolor' superlattices, in which CeCoIn$_5$ BLs are sandwiched by two different nonmagnetic metals, YbCoIn$_5$ and YbRhIn$_5$, as illustrated in Fig.\,\ref{tricolor_schematic}(b).  In these two types of Kondo superlattices,  the weakening of the Pauli paramagnetic pair breaking effect is more pronounced than that in `bicolor' CeCoIn$_5$/YbCoIn$_5$ superlattices, as revealed by the further enhancement of $H_{c2\perp}/H_{c2\perp}^{\rm orb}(0)$.  

In particular, in the tricolor Kondo superlattices, the Rashba spin-orbit- interaction induced global inversion symmetry breaking  is largely tunable by changing the layer thicknesses of  YbCoIn$_5$ and YbRhIn$_5$, leading  to profound changes in the superconducting properties of 2D CeCoIn$_5$ BLs.    Remarkably, the temperature  dependence of  $H_{c2\parallel}$ of YbCoIn$_5$(3)/CeCoIn$_5(n)$/YbRhIn$_5$(3), in which 3-UTC YbCoIn$_5$, $n$-UCT CeCoIn$_5$ ($n$ = 5 and 8) and 3-UCT YbRhIn$_5$ are stacked alternatively, in-plane upper critical field exhibits an anomalous upturn at low temperatures, which is attributed  to a possible emergence of a helical or stripe superconducting phase  \cite{Naritsuka_2017}. These results demonstrate that the tricolor  Kondo superlattices provide a new playground for exploring exotic superconducting states in the strongly correlated 2D electron systems with the Rashba effect.

The fabrication of tricolor superlattices containing $d$-wave superconducting layers offers the prospect of achieving even more fascinating pairing states than bulk CeCoIn$_5$, such as helical and stripe superconducting states \cite{Kaur_2005}, a pair-density-wave state \cite{Yoshida_2012}, complex stripe state \cite{Yoshida_2013}, a topological crystalline superconductivity \cite{Yoshida_2015,Watanabe_2015}, and Majorana fermion excitations\cite{Yoshida_2017,Daido_2016,Daido_2017,Sato_2009,Sato_2010}, in strongly correlated electron systems.

\section{Tuning the Pairing Interaction through the interface}
\subsection{CeCoIn$_5$/CeRhIn$_5$ Kondo superlattices}

\begin{figure}[htp]
	\centering
	\includegraphics[width=\linewidth]{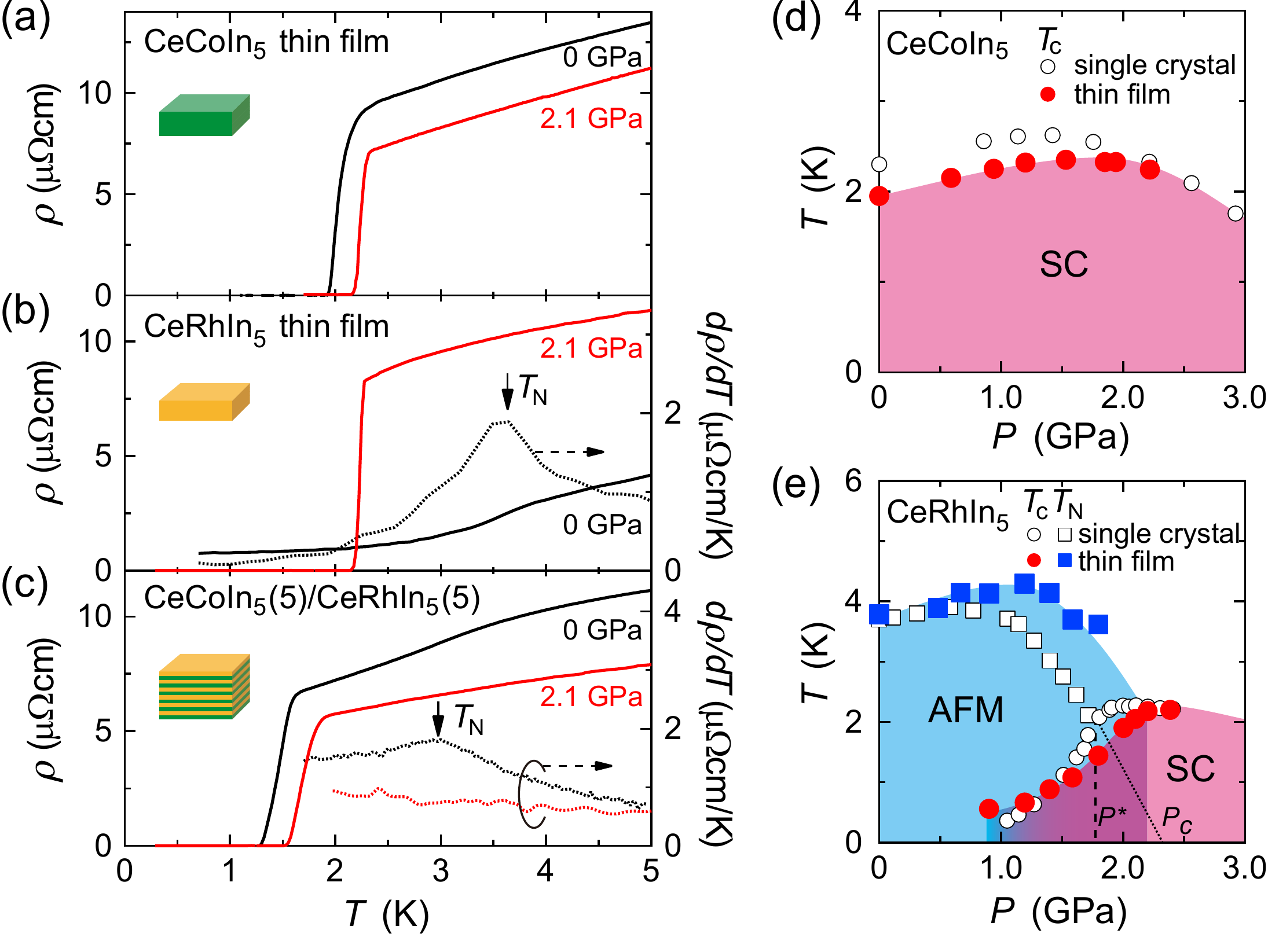}
	\caption{
		(a) Temperature dependence of the resistivity of CeCoIn$_5$ thin film at ambient pressure and at $P =$ 2.1 GPa. (b), (c) Temperature dependence of the resistivity (solid lines, left axes) and its temperature derivative $d\rho(T)/dT$ (dotted lines, right axes) for CeRhIn$_5$ thin film and CeCoIn$_5$(5)/CeRhIn$_5$(5) superlattice at ambient pressure and at $P =$ 2.1 GPa, respectively. The peak of $d\rho(T)/dT$ corresponds to AFM transition.
		(d), (e) $P-T$ phase diagrams of thin films and single crystals of (d) CeCoIn$_5$ and (e) CeRhIn$_5$.
	}
	\label{fig:res_thinfilm}
\end{figure}

Figures\,\ref{fig:res_thinfilm}(a), (b) and (c) depict the temperature dependence of the resistivity and its temperature derivative, $d\rho/dT$, for CeCoIn$_5$ and  CeRhIn$_5$ thin films and CeCoIn$_5$(5)/CeRhIn$_5$(5) superlattice at ambient and under pressure ($P=2.1$\,GPa), respectively.     The peak of $d\rho/dT$ corresponds to the AFM transition.  The pressure dependence of $T_c$ and $T_N$ for  CeCoIn$_5$ and CeRhIn$_5$ thin films, along with those for single crystals, are shown in Figs\,\ref{fig:res_thinfilm}(d) and \ref{fig:res_thinfilm}(e).   The $P-T$ phase diagrams of both films are essentially similar to those of single crystals.  However, $T_c$ (= 2.0\,K) in the CeCoIn$_5$ thin film  is slightly reduced from the bulk value, whereas $T_\mathrm{N}$ (= 3.7\,K) of CeRhIn$_5$ thin film is almost the same as that in a single crystal. With applying pressure, $T_c$ of the CeCoIn$_5$ thin film increases and shows a broad peak near $P \sim $ 1.7 GPa. Similar to CeRhIn$_5$ single crystals  \cite{Park_2008b,Knebel_2008}, superconductivity in the thin films develops at $P\gtrsim$ 1 GPa, where it coexists with magnetic order. In analogy to CeRhIn$_5$ single crystals, there appears to be a purely superconducting state at $P\gtrsim$ 2.1 GPa, which is a slightly higher pressure than that required to remove evidence for AFM order in single crystals.

Figure\,\ref{fig:hybrid_PTandHc2}(a) shows the $P$-dependence of $T_c$ and $T_\mathrm{N}$ determined by the peak in $d\rho(T)/dT$ for CeCoIn$_5$(5)/CeRhIn$_5$(5) superlattice. At $P \sim$ 2\,GPa, $T_c$ is at a maximum, forming a dome-shaped $P$-dependence. With pressure, $T_\mathrm{N}$ gradually decreases at low $P$ and decreases sharply when it exceeds $P\gtrsim$ 1\,GPa. At $P\gtrsim$ 1.6\,GPa, evidence of magnetic order is hidden beneath the superconducting dome. Although there is large ambiguity in determining a critical pressure $P_c$, a simple extrapolation of $T_\mathrm{N}$ gives $P_c \sim$ 2\,GPa, where $T_c$ has a maximum. Furthermore, this critical value is very close to the $P_c$ of CeRhIn$_5$ single crystal.

The $T_c$ and $T_\mathrm{N}$ of the hybrid superlattice are lower than those of the CeCoIn$_5$ and CeRhIn$_5$ thin films, suggesting that the reduction in dimensions affected the electronic structure. However, these values are higher than the corresponding CeCoIn$_5$/YbCoIn$_5$ and CeRhIn$_5$/YbRhIn$_5$, indicating the importance of interaction between CeCoIn$_5$ and CeRhIn$_5$ BLs.

\subsection{Superconductivity and antiferromagnetism in specially separated layers}
\begin{figure}[htp]
	\centering
	\includegraphics[width=0.8\linewidth]{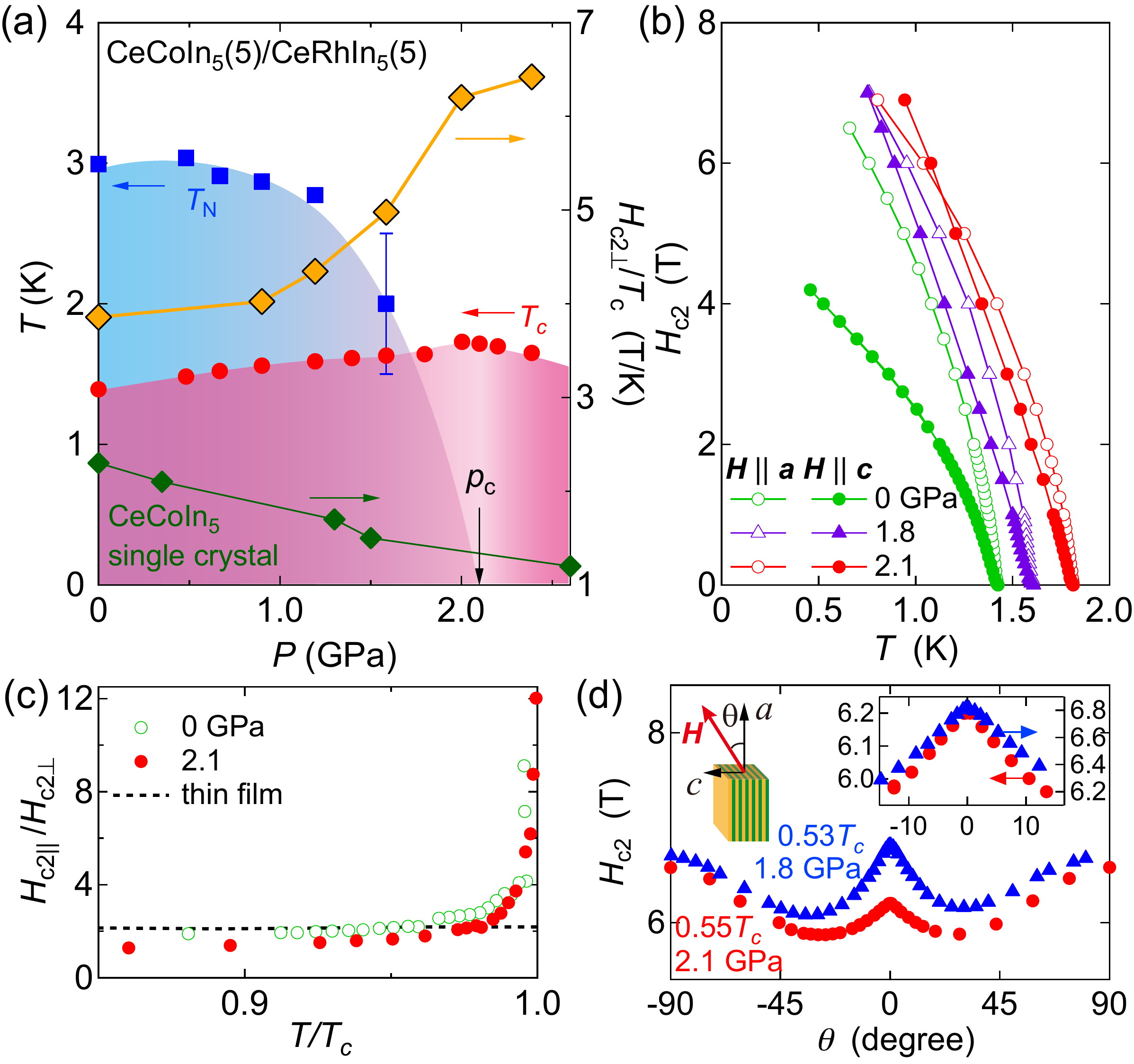}
	\caption{
		(a) $P-T$ phase diagram of CeCoIn$_5$(5)/CeRhIn$_5$(5) superlattice. Out-of-plane upper critical field $H_{c2\perp}$ normalized by $T_c$, $H_{c2\perp}/T_c$, measures the coupling strength of the superconductivity. (b) Temperature dependence of in-plane and out-of-plane upper critical fields at ambient pressure and at $P$ = 1.8 and 2.1\,GPa. (c) Anisotropy of upper critical field, $H_{c2\parallel}/H_{c2\perp}$, near $T_c$ of superlattices at ambient pressure and at 2.1\,GPa, along with the data of CeCoIn$_5$ thin film. (d) Angular dependence of upper critical field of superlattice at $P$ = 1.8 and 2.1\,GPa. (Inset) An expanded view of low angle region.
	}
	\label{fig:hybrid_PTandHc2}
\end{figure}

We show that 2D superconductivity is realized in the CeCoIn$_5$ BLs in the whole pressure regime in CeCoIn$_5$(5)/CeRhIn$_5$(5) superlattice. Figure\,\ref{fig:hybrid_PTandHc2}(b) depicts the $T$-dependence of the upper critical field determined by the midpoint of the resistive transition in a magnetic field $H$ applied parallel ($H_{c2\parallel}$) and perpendicular ($H_{c2\perp}$) to the $ab$ plane.  Figure\,\ref{fig:hybrid_PTandHc2}(c) shows the $T$-dependence of the anisotropy of upper critical fields, $H_{c2\parallel}/H_{c2\perp}$. Unlike the almost $T$-independent anisotropy seen in single crystals and thin films of CeCoIn$_5$, anisotropy in the superlattice shows a divergent increase toward $T_c$. This diverging anisotropy is characteristic of 2D superconductivity, in which $H_{c2\parallel}$ increases as $\sqrt{T_c-T}$ due to the Pauli paramagnetic limiting, but $H_{c2\perp}$ increases as $T_c-T$ due to orbital limiting near $T_c$. Considering this result and the fact that the 5-UCT of the CeCoIn$_5$ BL is comparable to the superconducting coherence length in the $c$-axis direction $\xi_{\perp} \sim 3$--4\,nm, the 5-UCT of the CeCoIn$_5$ BL effectively act as a 2D superconductor \cite{Mizukami_2011}. The 2D superconductivity is also confirmed from the angle variation of $H_{c2}(\theta)$. Figure \,\ref{fig:hybrid_PTandHc2}(d) and its inset show $H_{c2}(\theta)$ below and above $P^*$. For both pressures, at temperature well below $T_c$,
$H_{c2}(\theta)$ in the regime $|\theta|\lesssim30^{\circ}$ is enhanced with decreasing $|\theta|$ and exhibits a sharp cusp at $\theta = 0$. This cusp behavior is typical of Josephson-coupled layered superconductors \cite{Tinkham_1996}.

It should be noted  that in contrast to single crystals and thin films of CeRhIn$_5$, the CeRhIn$_5$ layers in CeCoIn$_5$/CeRhIn$_5$ hybrid superlattices do not become fully superconducting even under pressure where AFM order is suppressed. As a result, 2D superconductivity occurs in a wide pressure regime. In fact, as shown in Fig.\,\ref{fig:hybrid_PTandHc2}(d), at $P = 1.8$\,GPa where CeRhIn$_5$ thin film does not show bulk superconductivity, the hybrid superlattice shows an angular dependence with a cusp structure near $\theta = 0$. Essentially similar cusp-like behavior is observed at $P = 2.1$\,GPa above $P_c$, suggesting that 2D superconductivity derived from the CeCoIn$_5$ BLs is realized below and above $P_c$.

When the number of BL thickness is reduced, superconductivity survives in CeCoIn$_5$, but is suppressed in CeRhIn$_5$. This difference may be related to the ordering vector $\bm{q} = (0.5, 0.5, 0.297)$  \cite{Bao_2000} of the incommensurate magnetic structure of CeRhIn$_5$. In CeCoIn$_5$, on the other hand, the AFM fluctuations are dominated by $\bm{q} = (0.45, 0.45, 0.5)$  \cite{Raymond_2015}. This commensurability along the $c$-axis would match well with the superlattice structure, and as a result, the superconductivity is robust against the decrease in the BL thickness  \cite{Yamanaka_2015,Yamanaka_2017}.
Recent site-selective NMR measurements on CeCoIn$_5$/CeRhIn$_5$ superlattice have shown that AFM order is not induced in the CeCoIn$_5$ BLs \cite{Nakamine_2019}. The pressure suppresses magnetic order in CeRhIn$_5$ and CeCoIn$_5$ approaches the Fermi liquid state, so it is unlikely that AFM order is induced in the CeCoIn$_5$ BLs in the superlattice under pressure.
We comment on the reversal of $H_{c2}$ of the CeCoIn$_5$(5)/CeRhIn$_5$(5) superlattice at low temperature under pressure(Fig.\,\ref{fig:hybrid_PTandHc2}(b)). Such a reversed anisotropy of $H_{c2}$ can be seen in CeRhIn$_5$ single crystal in a high pressure region where AFM order is completely suppressed. However, similar reversed anisotropy ($H_{c2\perp}>H_{c2\parallel}$) is preserved at $P = 1.8$\,GPa, where $H_{c2\parallel}$ exceeds $H_{c2\perp}$ in CeRhIn$_5$ single crystal and thin film. This result suggests that the reversal of $H_{c2}$ occurs in 5-UCT CeCoIn$_5$ BLs. From the above results, we conclude that 2D superconductivity of CeCoIn$_5$ coupled by the Josephson effect within a BL is realized in the whole pressure regime.

Figure\,\ref{Fig5} displays $H_{c2\perp}(T)/H_{c2\perp}(0)$ of CeCoIn$_5$($n$)/CeRhIn$_5$($n$) and CeCoIn$_5$($n$)/YbCoIn$_5$(5) superlattices plotted as a function of $T/T_c$. Here $H_{c2\perp}^{\rm orb}(0)$ is calculated by the initial slope of $H_{c2\perp}(T)$ at $T_c$ by using Werthamer-Helfand-Hohenberg (WHH) formula, $H_{c2\perp}^{\rm orb}(0) = -0.69T_c(dH_{c2\perp}/dT)_{T_c}$. For comparison, we also include two extreme cases: $H_{c2\perp}/H^{orb}_{c2\perp}(0)$ for bulk CeCoIn$_5$ \cite{Tayama_2002}, in which $H_{c2}$ is dominated by Pauli paramagnetism, and the WHH curve with no Pauli effect. In CeCoIn$_5$($n$)/YbCoIn$_5$(5), $H_{c2\perp}/H_{c2\perp}^{\rm orb}(0)$ increases with decreasing $n$.
This is because the local inversion symmetry breaking suppresses the Pauli pair-breaking effect at the interfaces between BLs. As $n$ decreases, the contribution of the interface increases and the relative importance of orbital pair-breaking effect compared with Pauli pair-breaking effect increases. On the other hand, $H_{c2\perp}/H_{c2\perp}^{\rm orb}$ is almost independent on $n$ in CeCoIn$_5$($n$)/CeRhIn$_5$($n$), suggesting that the local inversion symmetry breaking in not important in the superlattices in which both substances constituting the superlattice are Ce-based compounds.

\subsection{Enhancement of superconducting pairing strength}
\begin{figure}[htp]\centering
	\includegraphics[width=\linewidth]{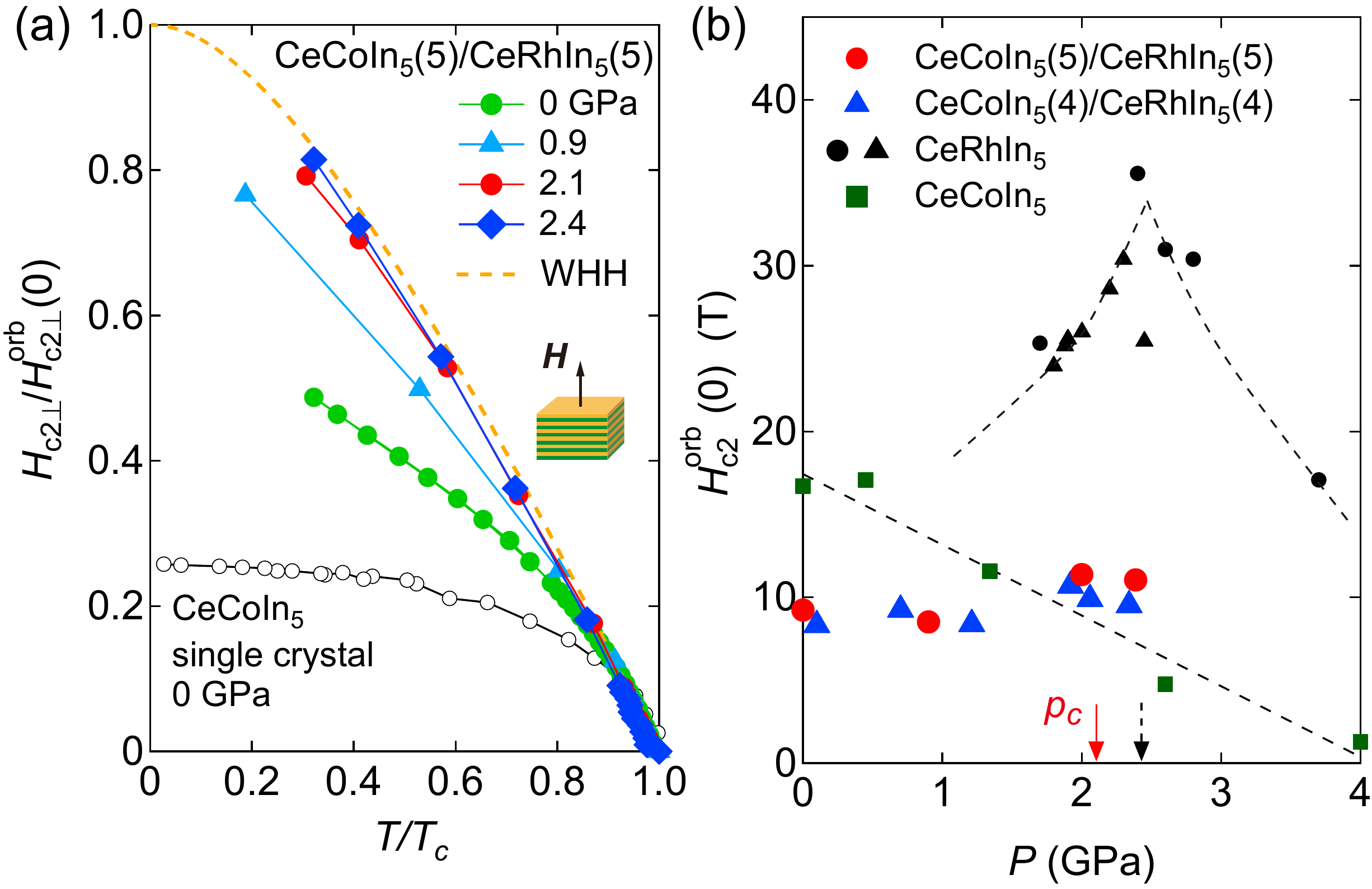}
	\caption{
		(a) Out-of-plane upper critical field $H_{c2\perp}$ normalized by the orbital-limited upper critical field at $T = 0$\,K, $H_{c2\perp}/H_{c2\perp}^{\rm orb}(0)$, for CeCoIn$_5$(5)/CeRhIn$_5$(5) superlattice is plotted as a function of the normalized temperature $T/T_c$. Two extreme cases, i.e., the result of the bulk CeCoIn$_5$ dominated by Pauli paramagnetic effect and the WHH curve with no Pauli effect, are also shown. (b) Pressure dependence of $H_{c2}^{\rm orb}(0)$ of CeCoIn$_5$($n$)/CeRhIn$_5$($n$) superlattices with $n = 4$ and 5 for ${\bm H}$$\parallel$$c$. For comparison, $H_{c2}^{\rm orb}(0)$ of CeRhIn$_5$ single crystals for ${\bm H}$$\parallel$$a$ and that of CeCoIn$_5$ single crystal for ${\bm H}$$\parallel$$c$ are shown.
		Solid and dashed arrows represent $P_c$ for CeCoIn$_5$($n$)/CeRhIn$_5$($n$) superlattices and CeRhIn$_5$ single crystal, respectively. 
	}
	\label{Fig6}
\end{figure}

The superconducting properties of the hybrid superlattice change dramatically when pressure is applied. Figure\,\ref{Fig6}(a) depicts the $T$-dependence of $H_{c2\perp}$/$H_{c2\perp}^{\rm orb}(0)$ of CeCoIn$_5$(5)/CeRhIn$_5$(5) for several pressures. Remarkably, near the critical pressure of $P_c \sim 2$\,GPa where AFM order vanishes, $H_{c2\perp}/H_{c2\perp}^{\rm orb}$ almost coincides with WHH curve, indicating that $H_{c2\perp}$ is determined only by the orbital pair-breaking effect.

The fact that $H_{c2\perp}$ reaches the orbital limit has important implications for the superconducting properties of the hybrid superlattice. In CeCoIn$_5$/YbCoIn$_5$, where YbCoIn$_5$ is a conventional metal, Pauli pair-breaking effect is weakened in the superlattice compared with the bulk due to local inversion symmetry breaking at the interfaces, where the Fermi surface splits with spin momentum locking due to anisotropic Rashba spin-orbit interaction. This leads to anisotropic suppression of the Zeeman effect which may be partly responsible for the observed reversed anisotropy $H_{c2\parallel}/H_{c2\perp}<1$ at low temperatures (Fig.\,\ref{fig:hybrid_PTandHc2}(d)). However, this effect is less important in CeCoIn$_5$($n$)/CeRhIn$_5$($n$) superlattices compared with CeCoIn$_5$/YbCoIn$_5$, which is evidenced by the fact that $H_{c2\perp}/H_{c2\perp}^{\rm orb}(0)$ does not strongly depend on $n$ (Fig.\,\ref{Fig5}). Furthermore, such an effect is not expected to show significant pressure dependence. Therefore, there must be a different mechanism that significantly enhances  $H_{c2\perp}^{\rm Pauli}$ given by Eq.\,(1).

An enhancement of $H_{c2\perp}^{\rm Pauli}$ is not due to a dramatic suppression of $g$. As $g$ is enhanced by pressure in both CeCoIn$_5$ and CeRhIn$_5$ \cite{Park_2008b}, $g$ is expected to be enhanced with pressure in the superlattice.
Therefore, a significant increase in the superconducting gap is thought to be the origin of the increase in $H_{c2\perp}^{\rm Pauli}$. This is also supported by the sharp increase in $H_{c2\perp}/T_c$ upon approaching $P_c$ shown in Fig.\,\ref{Fig6}(a).
Because $H_{c2\perp}\approx H_{c2\perp}^{\rm Pauli} \ll H_{c2\perp}^{\rm orb}(0)$ in the low $P$ regime and $H_{c2\perp}\approx H_{c2\perp}^{\rm orb}(0) \ll H_{c2\perp}^{\rm Pauli}$ near $P \sim p_c$, the enhancement of $H_{c2\perp}/k_\mathrm{B}T_c$ directly indicates an enhancement of $H_{c2\perp}^{\rm Pauli}/T_c$ and hence $\Delta/k_\mathrm{B}T_c$. This behavior is significantly different from CeCoIn$_5$ single crystal, in which $H_{c2\perp}/T_c$ monotonically decreases with pressure, approaching the Fermi liquid state. The enhancement of $\Delta/k_\mathrm{B}T_c$ is caused as a consequence of the enhancement of pairing interaction.
In the spin fluctuation mediated mechanism, the pairing interaction is brought about by high-energy spin fluctuations well above $\Delta$, while low-energy fluctuations cause the pair-breaking. High-energy fluctuations have the effect of increasing $T_c$, while low-energy fluctuations decrease $T_c$, so that the enhancement of pairing interaction can give rise to increase in $\Delta/k_\mathrm{B}T_c$ without accompanying a large enhancement of $T_c$. Therefore, these results demonstrate that the critical magnetic fluctuations developed in CeRhIn$_5$ BLs near its critical pressure are injected into CeCoIn$_5$ BLs through the interface and enhance the pairing interaction of the CeCoIn$_5$ BLs.

It has been established that normal and superconducting properties are greatly affected by quantum fluctuations in many classes of unconventional superconductors. The common behavior is that the effective mass of quasiparticle diverges as the system approaches a QCP, as reported in cuprates, pnictides and heavy-fermion systems  \cite{Shibauchi_2014,Shishido_2005,Ramshaw_2015}. Such an increase in effective mass gives rise to a corresponding enhancement $H_{c2}^{\rm orb}$, which is proportional to $(m^*\Delta)^2$. the CeCoIn$_5$/CeRhIn$_5$ superlattices show different behavior. In contrast to the CeRhIn$\_5$ single crystal, which shows a sharp peak at the critical pressure, the $H_{c2\perp}^{\rm orb}$ of the CeCoIn$_5$($n$)/CeRhIn$_5$($n$) superlattices with $n = 4$ and 5 does not show much $P$-dependent behavior, and there is no anomaly in $P_c$. Compared to a monotonic decrease of effective mass in CeCoIn$_5$ single crystal, the result of the hybrid superlattice is consistent with an enhancement of $\Delta$, indicating that there is no mass enhancement in CeCoIn$_5$ BLs. Such behavior is in contrast to what is expected for usual quantum criticality, and is a subject for future research.

\section{Coupling between superconductivity and antiferromagnetism}
To further examine how $d$-wave superconductors and antiferromagnets interact through the interface, we designed another hybrid superlattice using a different AFM metal CeIn$_3$ \cite{Naritsuka_2019}. The cubic CeIn$_3$ forms 3D AFM order with the ordered magnetic moment of 0.48\,$\mu_\mathrm{B}$ occurs with a commensurate wave vector $\bm{q}$ = (0.5, 0.5, 0.5) at $T_\mathrm{N}$ = 10\,K, where $\mu_\mathrm{B}$ is the Bohr magneton  \cite{Benoit_1980}. This is contrast to CeRhIn$_5$, which forms an incommensurate helical AFM order with $\bm{q}$ = (0.5, 0.5, 0.239) at $T_\mathrm{N}$ = 2.3\,K. On the other hand, both are Ce-based heavy-fermion AFM metal with AFM QCP under pressure \cite{Mathur_1998,Knebel_2001}. Therefore, it becomes possible to investigate the effect of different types of antiferromagnetism on $d$-wave superconductivity by measuring the $H_{c2}$ for CeCoIn$_5$/CeIn$_3$ superlattice under pressure, as have done for CeCoIn$_5$/CeRhIn$_5$.

\subsection{Robust magnetism against thickness reduction}
\begin{figure}[htp]
	\centering
	\includegraphics[width=\linewidth]{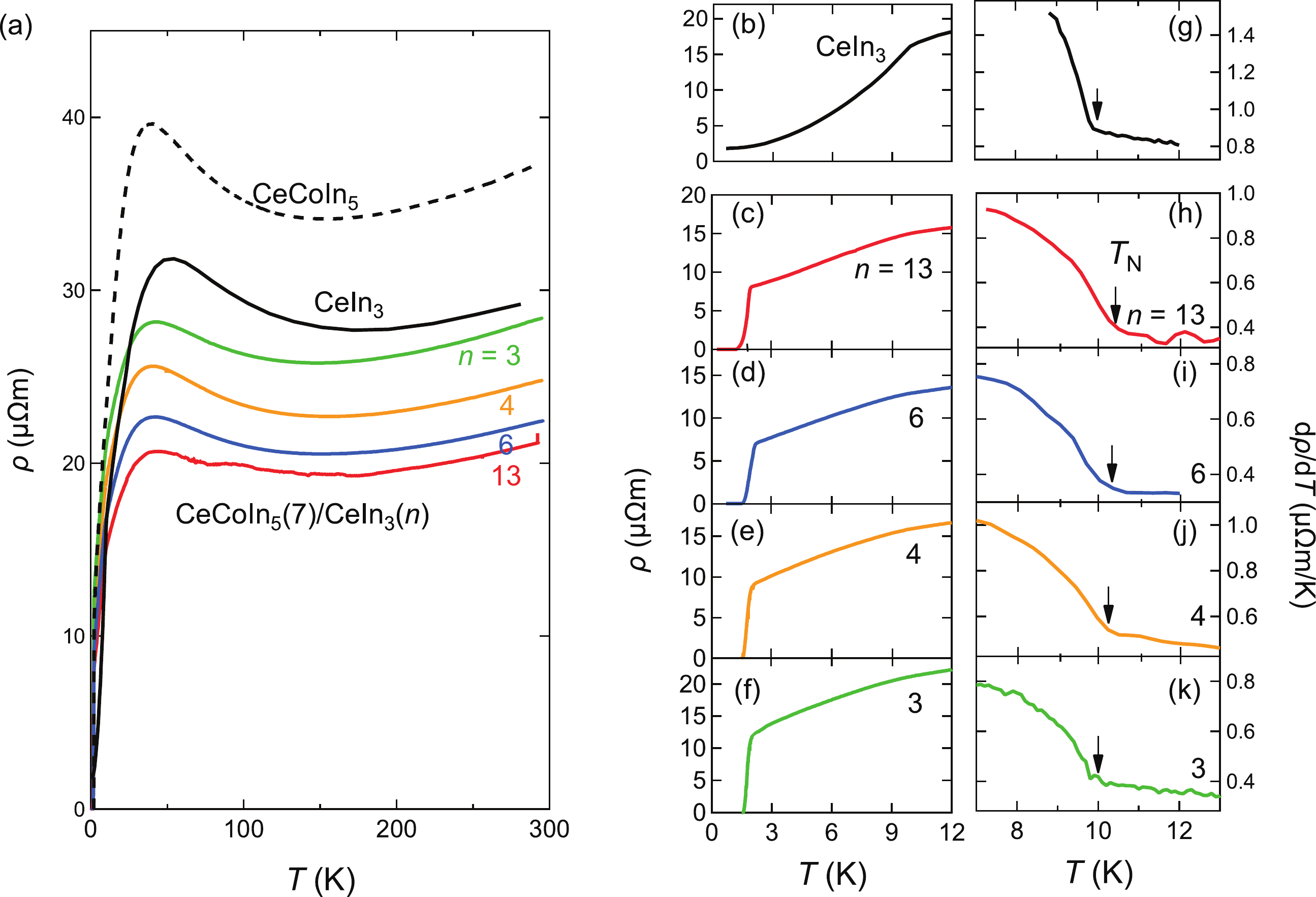}
	\caption{(a) Temperature dependence of the resistivity $\rho(T)$ in CeCoIn$_5$(7)/CeIn$_3$($n$) superlattices for $n$ = 3, 4, 6, and 13, along with $\rho(T)$ for CeIn$_3$ (black solid line) and CeCoIn$_5$ (black dashed line) thin films. Inset illustrates the schematics of CeCoIn$_5$(7)/CeIn$_3$($n$) superlattice. (b)-(f) $\rho(T)$ at low temperatures. (g)-(f) Temperature derivative of the resistivity, $d\rho(T)/dT$, as a function of temperature. The arrows indicate the N\'{e}el temperature $T_\mathrm{N}$. 
	}
	\label{Fig8}
\end{figure}

Figure\,\ref{Fig8}(a) depicts the temperature dependence of the resistivity $\rho$ of CeCoIn$_5$(7)/CeIn($n$) superlattices with $n$ = 3, 4, 6 and 13. We also show $\rho$ of CeCoIn$_5$ and CeIn$_3$ thin films grown by MBE. The mean free path of these superlattices is difficult to estimate because of the parallel conductions of CeCoIn$_5$ and CeIn$_3$ BLs. However, the mean free path in each BL is expected to be shorter than the atomically flat regions extending over distances of $\sim0.1~\mu$m, because of the following reasons. In CeCoIn$_5$ and CeIn$_3$ single crystals, the mean free path determined by the de Haas-van Alphen oscillations is $\sim 0.2\mu$m  \cite{Ebihara_1993,Harrison_2004}. The residual resistivity ratio of CeCoIn$_5$ and CeIn$_3$ thin films with 100\,nm thickness is 4--5 times smaller than that of the single crystals. Therefore, the mean free path of CeCoIn$_5$ and CeIn$_3$ BLs in the superlattices is expected to be much shorter than 0.1$~\mu$m, suggesting that the transport properties are not seriously influenced by the surface roughness.
The resistivity of CeCoIn$_5$(7)/CeIn($n$) superlattices follows the typical heavy-fermion behavior. With decreasing temperature, $\rho(T)$ increases below $\sim$ 150\,K due to the Kondo scattering but then begins to decrease due to strong $c$-$f$ hybridization between $f$-electrons and conduction ($c$) band electrons, leading to the narrow $f$-electron band at the Fermi level.
The Kondo coherence temperature $T_{coh}$, at which the formation of heavy-fermion occurs, is estimated from the maximum in $\rho(T)$. As shown in Fig.\,\ref{Fig8}(a), $T_{coh}$ of CeCoIn$_5$(7)/CeIn($n$) superlattices is nearly independent of $n$ and is closer to $T_{coh}$ of CeCoIn$_5$ thin film than $T_{coh}$ of CeIn$_3$ thin film, suggesting that $T_{coh}$ is mainly determined by CeCoIn$_5$ BLs.
Figure\,\ref{Fig8}(b)-(f) depict $\rho(T)$ at low temperatures. All superlattices show the superconducting transition at $T \approx$ 1.5\,K.
For the $n$ = 3- and 4-superlattices, $\rho(T)$ decreases with increasing slope, $d\rho(T)/dT$, as the temperature is lowered below 12\,K down to $T_c$.

The lattice parameters along the $a$-axis of CeCoIn$_5$, CeRhIn$_5$, and CeIn$_3$ is 4.613, 4.653, and 4.690 \AA, respectively. Therefore, a large tensile strain along the $a$-axis is expected in CeCoIn$_5$ BLs of CeCoIn$_5$/CeIn$_3$ compared to CeCoIn$_5$/CeRhIn$_5$. It has been shown that the uniaxial pressure dependence of $T_c$ along the $a$-axis for CeCoIn$_5$ is d$T_c$/d$P_a$ = 290 mK/GPa ( \cite{Oeschler_2003}), indicating that $T_c$ decreases by tensile strain. However, $T_c$ of CeCoIn$_5$/CeIn$_3$ ($T_c \sim$ 1.5 K) is larger than that of CeCoIn$_5$/CeRhIn$_5$ ($T_c \sim$ 1.4 K). We note that the lattice parameter along the $c$-axis for CeCoIn$_5$ BLs in CeCoIn$_5$/CeIn$_3$, which is estimated from x-ray diffraction, well coincides with that in CeCoIn$_5$/CeRhIn$_5$. These results suggest that the strain effect at the interfaces is not important for determining $T_c$. 
Figure\,\ref{Fig8}(g)--(k) display the temperature derivative of the resistivity $d\rho(T)/dT$. As shown by the arrows in Fig.\,\ref{Fig8}(g), $d\rho(T)/dT$ of CeIn$_3$ thin film exhibits a distinct kink at $T_\mathrm{N}$ = 10\,K  \cite{Benoit_1980}. Similar kink structures are observed in all superlattices at the temperatures indicated by arrows, showing the AFM transition.
\begin{figure}[htp]
	\centering
	\includegraphics[width=0.7\linewidth]{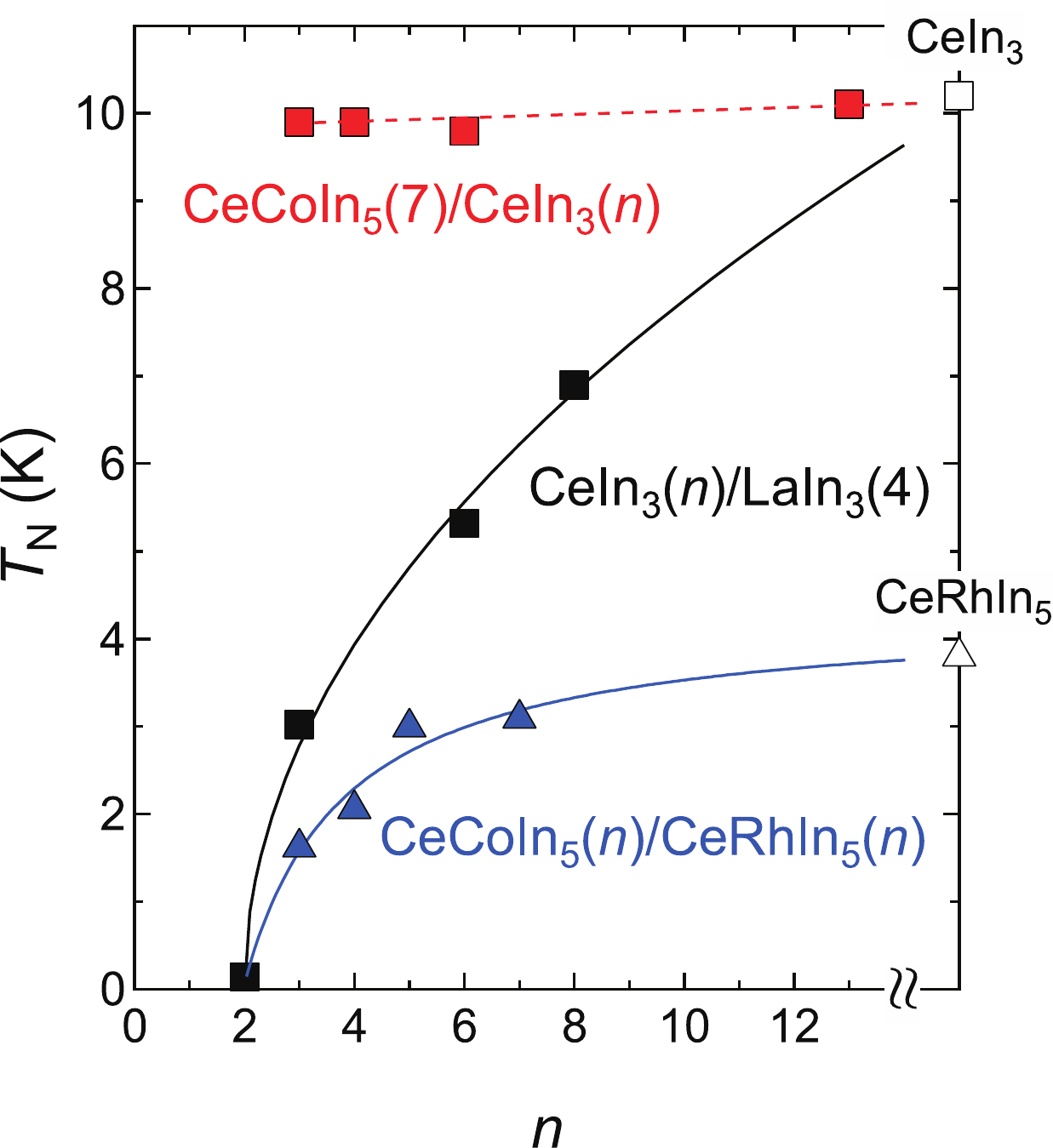}
	\caption{The N\'{e}el temperature $T_\mathrm{N}$ for CeCoIn$_5$(7)/CeIn$_3$($n$) as a function of $n$. For comparison, $T_\mathrm{N}$ for CeIn$_3$($n$)/LaIn$_3$(4) and CeCoIn$_5$($n$)/CeRhIn$_5$($n$) are shown. Open square and triangle are $T_\mathrm{N}$ of bulk CeIn$_3$ and CeRhIn$_5$ single crystals, respectively. 
	}
	\label{Fig9}
\end{figure}
Figure\,\ref{Fig9} shows the thickness dependence of $T_\mathrm{N}$ of the CeCoIn$_5$(7)/CeIn$_3$($n$) superlattices. For comparison, the data sets of CeIn$_3(4)$/LaIn$_3$($n$), where LaIn$_3$ is a nonmagnetic conventional metal with no $f$-electrons  \cite{Shishido_2010}, and CeCoIn$_5$($n$)/CeRhIn$_5$($n$) are also included in Fig.\,\ref{Fig9}.
Remarkably, the observed thickness dependence of $T_\mathrm{N}$ in CeCoIn$_5$/CeIn$_3$ is in striking contrast to that in CeIn$_3$/LaIn$_3$; While $T_\mathrm{N}$ is strongly suppressed with decreasing $n$ and vanishes at $n$ = 2 in CeIn$_3$/LaIn$_3$, $T_\mathrm{N}$ is nearly independent of $n$ in CeCoIn$_5$(7)/CeIn$_3$($n$).
This suggests that CeIn$_3$ BLs are coupled weakly by the RKKY interactions through the adjacent LaIn$_3$ BL, but they can strongly couple through the adjacent CeCoIn$_5$ BL. This is even more surprising, as the distance between different CeIn$_3$ BLs is larger in the CeCoIn$_5$(7)/CeIn$_3$($n$) superlattices than in the CeIn$_3$($n$)/LaIn$_3$(4) superlattices.
We thus conclude that small but finite magnetic moments are induced in CeCoIn$_5$ BLs in CeCoIn$_5$/CeIn$_3$, which mediate the RKKY-interaction. On the other hand, because of the absence of strongly interacting $f$-electrons in LaIn$_3$, which can form magnetic moments, the RKKY interaction in CeIn$_3$/LaIn$_3$ can be expected to be much weaker.
To clarify this, a microscopic probe of magnetism, such as NMR measurements, is required. We note that as shown in Fig.\,\ref{Fig9}, the reduction of $T_\mathrm{N}$ is also observed in CeCoIn$_5$($n$)/CeRhIn$_5$($n$) superlattices  \cite{Naritsuka_2018}, suggesting that the RKKY interaction between CeRhIn$_5$ BLs through adjacent CeCoIn$_5$ BL is negligibly small.
This is supported by the recent site-selective NMR measurements which report no discernible magnetic moments induced in the CeCoIn$_5$ BLs while magnetic fluctuations are injected from CeRhIn$_5$ BLs into one or two layers of CeCoIn$_5$ BLs in CeCoIn$_5$/CeRhIn$_5$  \cite{Nakamine_2019}.

\subsection{Tuning AFM fluctuations via pressure}
\begin{figure}[htp]
	\centering
	\includegraphics[width=\linewidth]{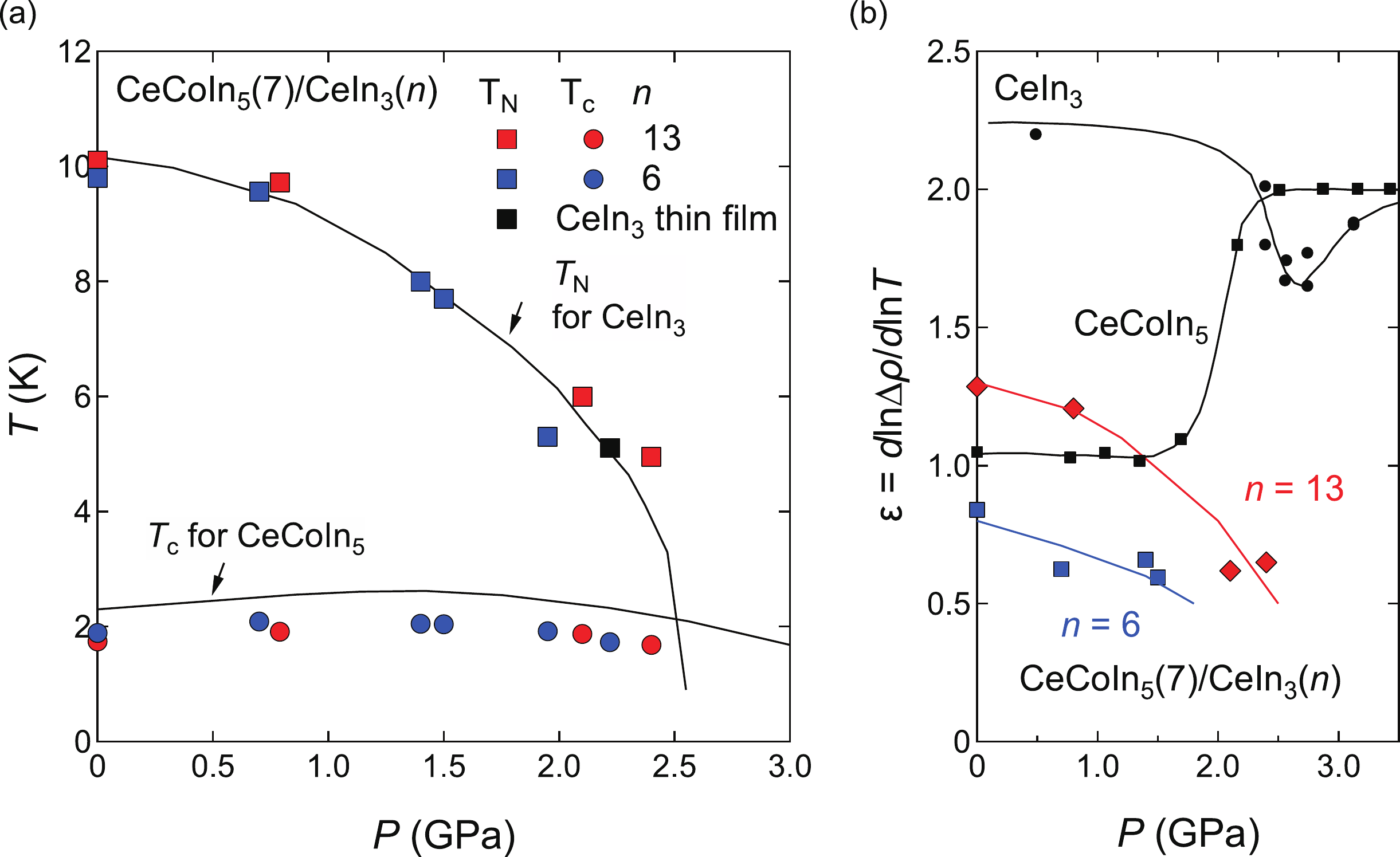}
	\caption{(a) Pressure dependence of $T_\mathrm{N}$ and $T_c$ of CeCoIn$_5$(7)/CeIn$_3$($n$) superlattices for $n$ = 13 and 6. For comparison, $T_\mathrm{N}$ of CeIn$_3$ and $T_c$ of CeCoIn$_5$ single crystals are shown by solid lines. (b) Pressure dependence of the exponent $\varepsilon$ in $\rho(T) = \rho_0+AT^\varepsilon$, obtained from $d\ln\Delta\rho/d\ln T$ ($\Delta\rho = \rho(T)-\rho_0$), for the CeCoIn$_5$(7)/CeIn$_3$($n$) superlattices for $n$ = 13 and 6. For comparison, $\varepsilon$ for bulk CeIn$_3$ and CeCoIn$_5$ single crystals is shown. 
	}
	\label{Fig11}
\end{figure}

The pressure dependence of the superconducting and magnetic properties provide crucial information on the mutual interaction between superconductivity and magnetism through the interface.
Figure\,\ref{Fig11}(a) depicts the pressure dependence of $T_\mathrm{N}$ and $T_c$ for CeCoIn$_5$(7)/CeIn$_3$($n$) superlattices for $n$ = 6 and 13. With applying pressure, $T_\mathrm{N}$ decreases rapidly. For comparison, $T_\mathrm{N}$ of a single crystal CeIn$_3$ is also shown by the solid line  \cite{Mathur_1998}. The pressure dependence of $T_\mathrm{N}$ of both superlattices is very similar to that of the bulk CeIn$_3$ single crystal. In bulk CeIn$_3$ crystal, the AFM QCP is located at $P_c \approx 2.6$\,GPa. It is natural to expect, therefore, that the AFM QCP of the superlattices is close to 2.6\,GPa. Thus, at 2.4\,GPa, the superlattices are in the vicinity of the AFM QCP.  This is supported by the temperature dependence of the resistivity under pressure. The resistivity can be fitted as
\begin{equation}
\rho(T) = \rho_0+AT^{\varepsilon}.
\end{equation}
Figure\,\ref{Fig11}(b) shows the pressure dependence of $\varepsilon$ obtained from $d \ln\Delta \rho/d \ln T$, where $\Delta\rho = \rho(T)-\rho_0$.  The magnitude of $\varepsilon$ decreases with pressure. In bulk CeIn$_3$ single crystal, $\varepsilon$ decreases with pressure and exhibits a minimum at the AFM QCP \cite{Mathur_1998,Knebel_2001}.
On the other hand, applying pressure to CeCoIn$_5$ leads to an increase of $\varepsilon$, which is attributed to the
suppression of the non-Fermi liquid behavior, $\rho(T)\propto T$, and the development of a Fermi liquid state with its characteristic $\rho(T)\propto T^2$ dependence \cite{Sidorov_2002,Nakajima_2007}.
Therefore, the reduction of $\varepsilon$ with pressure arises from the CeIn$_3$ BLs, indicating that the CeIn$_3$ BLs approach the AFM QCP.

\begin{figure}[htp]
	\centering
	\includegraphics[width=0.6\linewidth]{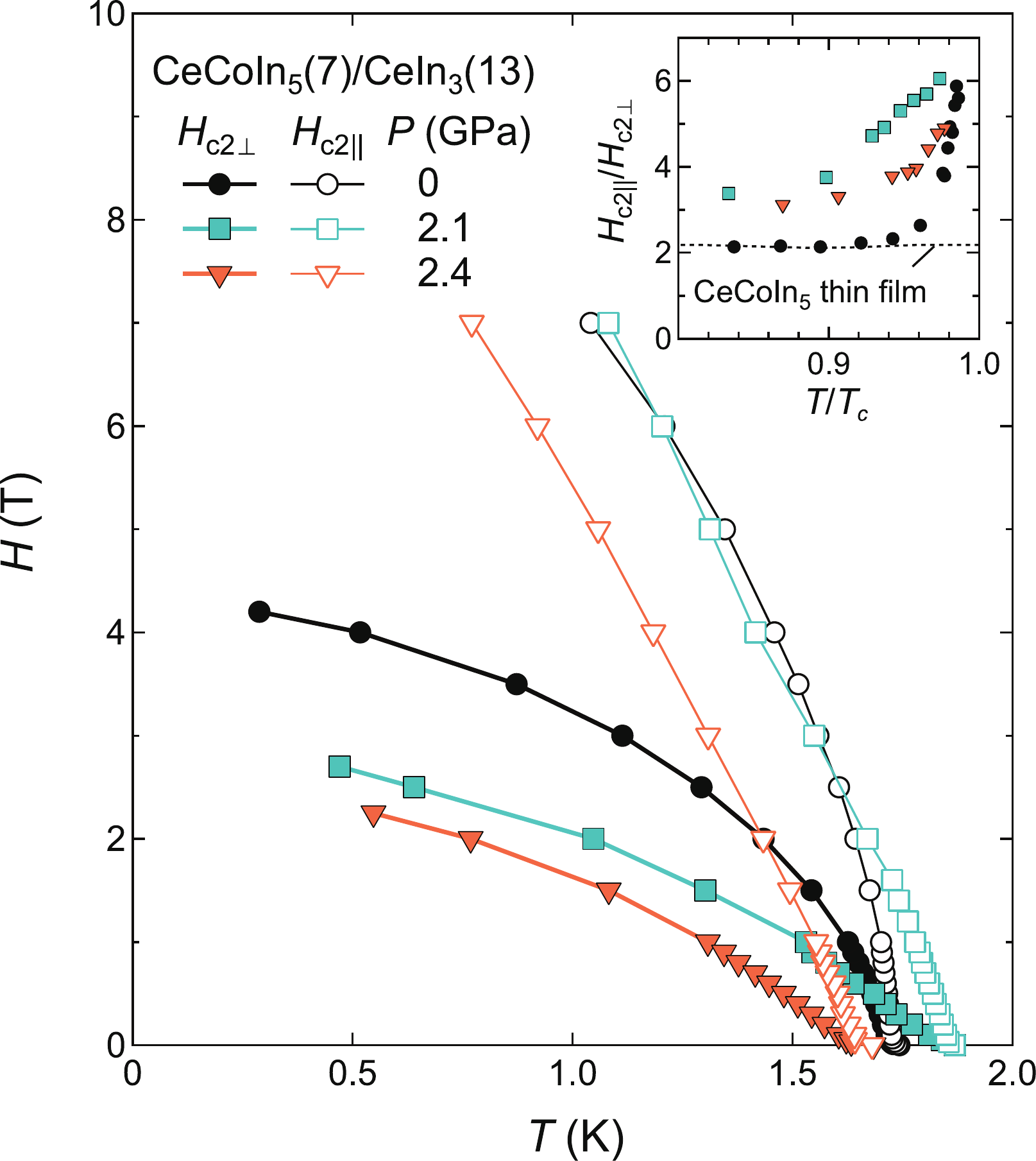}
	\caption{Temperature dependence of upper critical fields in magnetic fields parallel ($H_{c2\parallel}$, open symbols) and perpendicular ($H_{c2\perp}$, closed symbols) to the $ab$-plane for CeCoIn$_5$(7)/CeIn$_3$(13) superlattice at ambient pressure and at 2.1 and 2.4\,GPa. The inset shows anisotropy of the upper critical field, $H_{c2\parallel}/H_{c2\perp}$. The data of CeCoIn$_5$ thin film at ambient pressure is shown by dotted line. 
	}
	\label{Fig12}
\end{figure}

As shown in Fig\,\ref{Fig11}(a), $T_c$ increases, peaks at $\sim$ 1.8\,GPa, and then decreases when applying pressure. This pressure dependence bears a resemblance to that of CeCoIn$_5$ bulk single crystals  \cite{Sidorov_2002}.
An analysis of the upper critical field provides important information about the superconductivity of CeCoIn$_5$ BLs. Figure\,\ref{Fig12} depicts the temperature dependence of the upper critical field determined by the midpoint of the resistive transition in a magnetic field $\bm{H}$ applied parallel ($H_{c2\parallel}$) and perpendicular ($H_{c2\perp}$) to the layers. The inset of Fig\,\ref{Fig12} shows the anisotropy of the upper critical fields $H_{c2\parallel}/H_{c2\perp}$ at ambient pressure. The 2D feature is revealed by the diverging anisotropy of $H_{c2\parallel}/H_{c2\perp}$ of the superlattice on approaching $T_c$, in sharp contrast to the CeCoIn$_5$ thin film. Thus, each CeCoIn$_5$ BL in CeCoIn$_5$/CeIn$_3$ superlattice effectively behaves as a 2D superconductor.

\subsection{Effect of magnetic fluctuations on superconductivity}
\begin{figure}[htp]
	\centering
	\includegraphics[width=\linewidth]{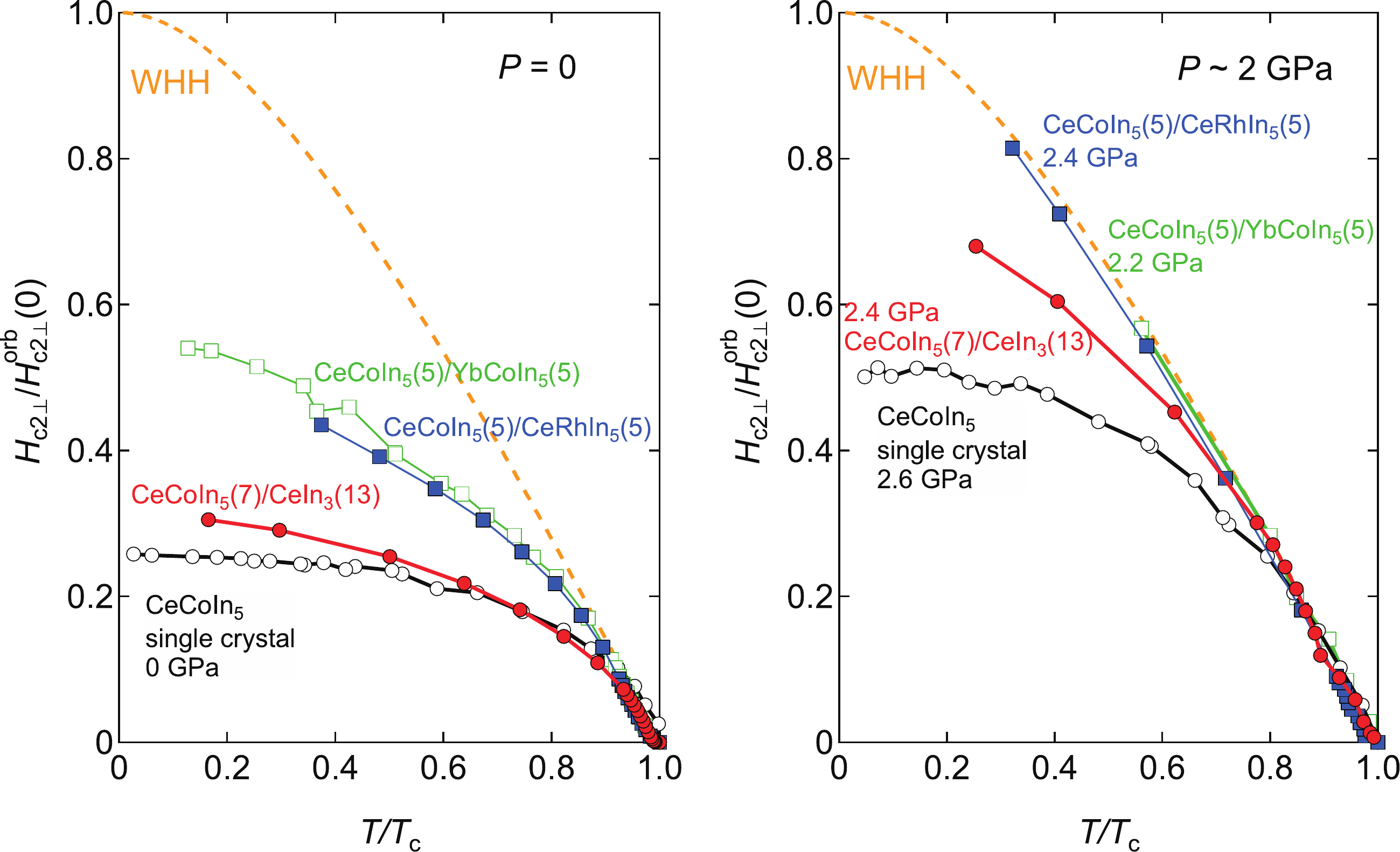}
	\caption{(a) Upper critical field in perpendicular field normalized by the orbital limiting upper critical field, $H_{c2\perp}/H_{c2\perp}^\mathrm{orb}(0)$, plotted as a function of $T/T_c$ (a) at ambient pressure and (b) under pressure about 2\,GPa for CeCoIn$_5$(7)/CeIn$_3$(13) superlattices. For comparison, $H_{c2\perp}/H_{c2\perp}^\mathrm{orb}(0)$ for bulk CeCoIn$_5$ single crystal, CeCoIn$_5$(5)/YbCoIn$_5$(5) and CeCoIn$_5$(5)/CeRhIn$_5$(5) are shown. Orange dotted lines represent the WHH curve, which is upper critical field for purely orbital limiting. 
	}
	\label{Fig13}
\end{figure}

It has been revealed that the $T$ dependence of $H_{c2\perp}$ provides crucial information on the effect of interfaces on the superconductivity of CeCoIn$_5$ BLs.
In particular, the modification of the Pauli paramagnetic effect in the superlattice, which dominates the pair breaking in bulk CeCoIn$_5$ single crystals, provide valuable clues \cite{Goh_2012,Shimozawa_2014,Naritsuka_2017,Naritsuka_2018}.
Figure\,\ref{Fig13}(a) and (b) depict the $T$ dependence of the $H_{c2\perp}$ of CeCoIn$_5$(7)/CeIn$_3$(13) superlattice at ambient pressure Fig.\,\ref{Fig13}(a) and under pressure Fig.\,\ref{Fig13}(b), normalized by the orbital-limited upper critical field at zero temperature, $H_{c2\perp}^\mathrm{orb}(0)$, which is obtained from the WHH formula, $H_{c2\perp}^\mathrm{orb}(0) = -0.69T_c(dH_{c2\perp}/dT)_{T_c}$ \cite{Werthamer_1966}. In figure\,\ref{Fig13}(a) and \ref{Fig13}(b), two extreme cases are also included; the WHH curve with no Pauli pair-breaking and $H_{c2}/H_{c2\perp}^\mathrm{orb}(0)$ for bulk CeCoIn$_5$ single crystal  \cite{Tayama_2002}.  For comparison, $H_{c2\perp}^\mathrm{orb}(0)$ for CeCoIn$_5$/YbCoIn$_5$ and CeCoIn$_5$/CeRhIn$_5$ are also shown \cite{Mizukami_2011, Naritsuka_2018}.

At ambient pressure, $H_{c2\perp}/H_{c2\perp}^\mathrm{orb}(0)$ is significantly increased from bulk CeCoIn$_5$ single crystal in both CeCoIn$_5$/YbCoIn$_5$ and CeCoIn$_5$/CeRhIn$_5$, indicating the suppression of the Pauli paramagnetic pair-breaking effect. However, we point out that the mechanisms of this suppression in these two systems are essentially different. In CeCoIn$_5$/YbCoIn$_5$, as discussed in section IV-B,  the enhancement of $H_{c2\perp}/H_{c2\perp}^\mathrm{orb}(0)$ is caused by the local ISB at the interface  \cite{Goh_2012, Maruyama_2012}. 
At $P$ = 2.2\,GPa, $H_{c2\perp}/H_{c2\perp}^\mathrm{orb}(0)$ of CeCoIn$_5$/YbCoIn$_5$ nearly coincides with the WHH curve, indicating that $H_{c2\perp}$ is dominated by the orbital pair breaking most likely due to the suppression of the Pauli paramagnetic pair-breaking effect by the Rashba splitting.

As shown in Fig.\,\ref{Fig5},  in stark contrast to  CeCoIn$_5(n)$/YbCoIn$_5(n)$, where  $H_{c2\perp}/H_{c2\perp}^\mathrm{orb}(0)$ is strongly enhanced with decreasing $n$, $H_{c2\perp}/H_{c2\perp}^\mathrm{orb}(0)$  in CeCoIn$_5(n)$/CeRhIn$_5(n)$ is independent of $n$.  This indicates  that the effect of the local ISB on $H_{c2\perp}$ is much less important  in CeCoIn$_5$/CeRhIn$_5$ than CeCoIn$_5$/YbCoIn$_5$, possibly owing to smaller asymmetric  potential gradient at Ce/Ce interface compared with that at Ce/Yb one  \cite{Naritsuka_2018}.  
It has been proposed that magnetic fluctuations in CeRhIn$_5$ BLs injected through the interface dramatically enhance the pairing interaction in CeCoIn$_5$ BLs, leading to the enhancement of $\Delta$.  As a result, $H_{c2\perp}^\mathrm{Pauli}$ is enhanced. This raises the relative importance of the orbital pair-breaking effect, giving rise to the enhancement of $H_{c2\perp}/H_{c2\perp}^\mathrm{orb}(0)$  \cite{Naritsuka_2018}. At $P$ = 2.1\,GPa, which is close to the AFM QCP of CeRhIn$_5$ BLs, $H_{c2\perp}/H_{c2\perp}^\mathrm{orb}(0)$ nearly coincides with the WHH curve. This has been attributed to the enhanced Pauli limiting field that well exceeds the orbital limiting field ($H_{c2\perp}^\mathrm{Pauli}\gg H_{c2\perp}^\mathrm{orb}$).

In contrast to CeCoIn$_5$/YbCoIn$_5$ and CeCoIn$_5$/CeRhIn$_5$, $H_{c2\perp}/H_{c2\perp}^\mathrm{orb}(0)$ only show a slight increase from CeCoIn$_5$(7)/CeIn$_3$(13) superlattice at ambient pressure from that of bulk CeCoIn$_5$ single crystal. This indicates that $H_{c2\perp}$ is dominated by Pauli paramagnetic effect, i.e. $H_{c2\perp} \approx H_{c2\perp}^\mathrm{Pauli}\ll H_{c2\perp}^\mathrm{orb}$. 
This implies that the effect of local inversion symmetry breaking on the superconductivity in CeCoIn$_5$/CeIn$_3$ is weak compared with CeCoIn$_5$/YbCoIn$_5$.  The local inversion symmetry is broken for the CeCoIn$_5$/YbCoIn$_5$ on the CoIn-layer while it is broken on the Ce layer for CeCoIn$_5$/CeIn$_3$ and CeCoIn$_5$/CeRhIn$_5$. Therefore, the present results suggest that the inversion symmetry breaking on the CoIn-layer induces a larger local electric field gradient.  Moreover, superconducting electrons in CeCoIn$_5$ BLs are not strongly influenced by the AFM order in CeIn$_3$ BLs compared with CeCoIn$_5$/CeRhIn$_5$.

\begin{figure}[htp]
	\centering
	\includegraphics[width=0.6\linewidth]{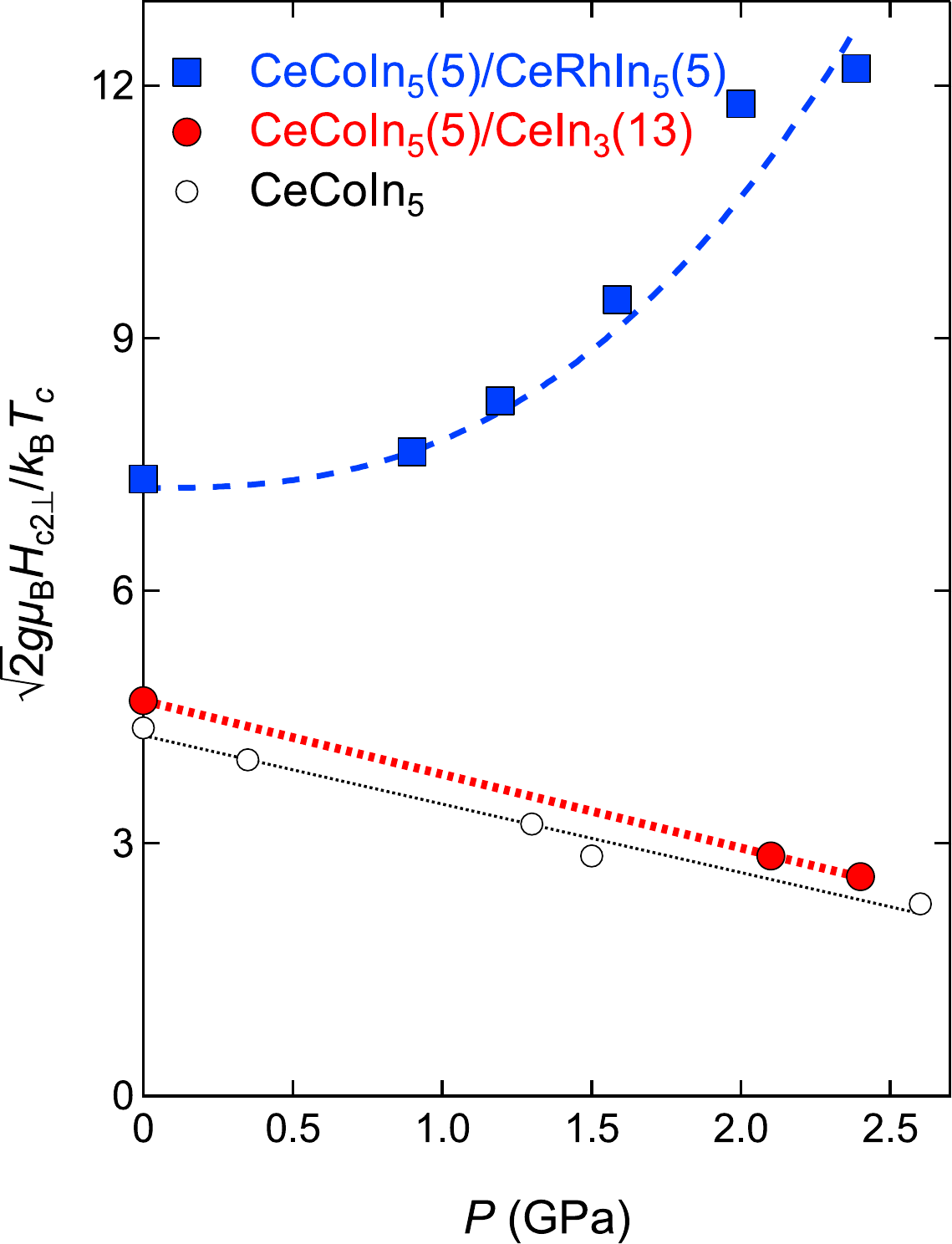}
	\caption{Pressure dependence of $q = \sqrt{2}g\mu_\mathrm{B}H_{c2\perp}/k_\mathrm{B}T_c \approx 2\Delta/k_\mathrm{B}T_c$ for CeCoIn$_5$(7)/CeIn$_3$(13) superlattice. For comparison, $q$ of bulk CeCoIn$_5$ single crystal and CeCoIn$_5$(5)/CeRhIn$_5$(5) are plotted. 
	}
	\label{Fig14}
\end{figure}

When superconductivity is dominated by the Pauli-limiting effect ($H_{c2\perp} \approx H_{c2\perp}^\mathrm{Pauli}$), $2\Delta/k_\mathrm{B}T_c$ is obtained from Eq.\ (\ref{eq:Pauli_limiting_field}) as
\begin{equation}
\frac{2\Delta}{k_\mathrm{B} T_c} \approx \sqrt{2} \frac{g\mu_\mathrm{B}H_{c2\perp}}{k_\mathrm{B}T_c}.
\end{equation}
where $\mu_\mathrm{B}$ is Bohr magneton and $g$ is the $g$-factor of electron. In Fig.\,\ref{Fig14}, $g$ = 2 is assumed.
Figure \ref{Fig14} depicts the pressure dependence of $q = \sqrt{2}g\mu_\mathrm{B}H_{c2\perp}/k_\mathrm{B}T_c$ for CeCoIn$_5$/CeRhIn$_5$ and CeCoIn$_5$/CeIn$_3$, along with $q$ for bulk CeCoIn$_5$ single crystal. Here $g$ = 2 is assumed.
Although this simple assumption should be scrutinized, the fact that $q = 4.2$ of the bulk CeCoIn$_5$ is larger than the BCS value of $q$ = 3.54 is consistent with the strong coupling superconductivity, which is supported by the specific heat measurements that report $2\Delta/k_\mathrm{B}T_c \approx 6$  \cite{Petrovic_2001b}.
The increase of $q$ with pressure in CeCoIn$_5$/CeRhIn$_5$ implies the increase of $2\Delta/k_\mathrm{B}T_c$. This increase has been attributed to an enhancement of the force binding superconducting electron pairs. In this case, an increase of $2\Delta/k_\mathrm{B}T_c$ occurs without accompanying a large enhancement of $T_c$, which is consistent with the results of CeCoIn$_5$/CeRhIn$_5$  \cite{Naritsuka_2018}. Thus, the critical AFM fluctuations that develop in CeRhIn$_5$ BLs near the QCP are injected into the CeCoIn$_5$ BLs through the interface and strongly enhance the pairing interaction in CeCoIn$_5$ BLs.

In stark contrast to CeCoIn$_5$/CeRhIn$_5$ superlattices, $q$ decreases with pressure in bulk CeCoIn$_5$ single crystal. This implies that the pairing interaction is weakened by applying pressure, which is consistent with the fact that the pressure moves the system away from the QCP of CeCoIn$_5$. The reduction of $2\Delta/k_\mathrm{B}T_c$ with pressure in bulk CeCoIn$_5$ single crystals is confirmed by the jump of the specific heat at $T_c$  \cite{Knebel_2004}. It should be stressed that the pressure dependence of $q$ in CeCoIn$_5$(7)/CeIn$_3$(13) is very similar to that of bulk CeCoIn$_5$. This strongly indicates that the pairing interactions in CeCoIn$_5$ BLs are barely influenced by AFM fluctuations injected from the adjacent CeIn$_3$ BLs through the interface even when CeIn$_3$ BLs are located near the AFM QCP.

The most salient feature in the CeCoIn$_5$/CeIn$_3$ superlattices is that the superconductivity of CeCoIn$_5$ BLs is little affected by the critical AFM fluctuations in CeIn$_3$ BLs, despite the fact that AFM fluctuations are injected from the adjacent CeIn$_3$ BLs into CeCoIn$_5$ BLs, as evidenced by the AFM order in CeCoIn$_5$/CeIn$_3$ demonstrating that different CeIn$_3$ BLs are magnetically coupled by the RKKY interaction through adjacent CeCoIn$_5$ BLs.
Even in the vicinity of the AFM QCP of the CeIn$_3$ BLs, the superconducting state in the CeCoIn$_5$ BLs is very similar to that of CeCoIn$_5$ bulk single crystals. This indicates that the AFM fluctuations injected from CeIn$_3$ BLs do not help to enhance the force binding the superconducting electron pairs in CeCoIn$_5$ BLs. This is in stark contrast to CeCoIn$_5$/CeRhIn$_5$, in which the pairing force in CeCoIn$_5$ BL is strongly enhanced by the AFM fluctuations in CeRhIn$_5$ BLs \cite{Naritsuka_2018}, although the CeRhIn$_5$ BLs are magnetically only weakly coupled through CeCoIn$_5$ BLs.

We note that the superconducting phase appears under pressure in CeRhIn$_5$ single crystals and epitaxial thin films. On the other hand, in the CeRhIn$_5$/YbRhIn$_5$, zero resistivity is not attained even under pressure \cite{Naritsuka_2018}. This result indicates that the superconductivity of CeRhIn$_5$ is suppressed when the thickness of the BLs was reduced. Similarly, in CeCoIn$_5$/CeRhIn$_5$ superlattices, 2D superconductivity derived from the CeCoIn$_5$ BLs is thought to be realized from ambient pressure to under pressure near QCP.

\subsection {Contrasting behaviors between CeCoIn$_5$/CeIn$_3$ and CeCoIn$_5$/CeRhIn$_5$ superlattices}

As discussed in the previous sections, CeCoIn$_5$/CeIn$_3$ and CeCoIn$_5$/CeRhIn$_5$ superlattices exhibit contrasting superconducting and magnetic properties.  We point out that   there are two possible important factors that determine whether magnetic fluctuations are injected through the interface; One is the magnetic wave vector and the other is the matching of the Fermi surface between two materials.

For CeCoIn$_5$, the Fermi surface is 2D-like and AFM fluctuations with wave vector $\bm{q}_0$ = (0.45, 0.45, 0.5) are dominant \cite{Raymond_2015}. The magnetic wave vector in the ordered phase of CeIn$_3$ is commensurate with $\bm{q}_0$ = (0.5, 0.5, 0.5) \cite{Benoit_1980}. The evolution of the ordered moment below $T_\mathrm{N}$ is consistent with mean field theory. While the wave number along the $c$ axis, $q_c$, of CeIn$_3$ is the same as that of CeCoIn$_5$, the 3D Fermi surface of CeIn$_3$ is very different from the 2D Fermi surface of CeCoIn$_5$.
On the other hand, for CeRhIn$_5$, $\bm{q}_0$ in the ordered phase is incommensurate $\bm{q}_0$ = (0.5, 0.5, 0.297) at low pressure  \cite{Bao_2000} and changes to  $\bm{q}_0$ = (0.5, 0.5, 0.4) above $\sim 1.0$\,GPa  \cite{Raymond_2008}. Thus, the $q_c$ of CeRhIn$_5$ is different from the $q_c$ of CeCoIn$_5$.
The evolution of the ordered moment below $T_\mathrm{N}$ deviates from mean field behavior, likely due to 2D fluctuations.  However, the 2D Fermi surface of CeRhIn$_5$ bears a close resemblance to that of CeCoIn$_5$.

The equality between the $c$ axis component of $\bm{q}_0$ in
CeCoIn$_5$ and CeIn$_3$ would explain why the magnetic coupling between CeIn$_3$ BLs through a CeCoIn$_5$ BL is stronger than that between CeRhIn$_5$ BLs. Thus, AFM order is formed in CeCoIn$_5$(7)/CeIn$_3$($n$) even for small $n$, for which the AFM order has already vanished in CeCoIn$_5$($n$)/CeRhIn$_5$($n$).
In magnetically mediated superconductors, the pairing interaction is expected to be strongly wave number dependent. Considering the good resemblance of the Fermi surface and the same $d_{x^2-y^2}$ superconducting gap symmetry of CeCoIn$_5$ and CeRhIn$_5$ \cite{Park_2008b}, it is likely that the pairing interaction in both compounds has 2D character and peaks around the same wave number on the Fermi surface. Furthermore, it has been assumed that 2D magnetic fluctuations are strong in CeRhIn$_5$. Thus, superconductivity in the CeCoIn$_5$ BLs of CeCoIn$_5$($n$)/CeRhIn$_5$($n$) is strongly influenced.
On the other hand, in CeIn$_3$ with 3D Fermi surface, 2D AFM fluctuations are expected to be very weak. AFM fluctuations having 3D character in CeIn$_3$ may not play an important role for the pairing interaction in CeCoIn$_5$, resulting in little change of the superconductivity in CeCoIn$_5$/CeIn$_3$.

\section{Conclusion}
We reviewed the most recent advances of Kondo superlattices containing atomic layers of strongly correlated heavy fermion superconductor CeCoIn$_5$ with  $d$-wave symmetry, CeCoIn$_5$/YbCoIn$_5$, CeCoIn$_5$/CeRhIn$_5$  and CeCoIn$_5$/CeIn$_3$ grown by using a state-of-the-art MBE technique.   In these Kondo superlattices, superconducting heavy electrons are confined within the 2D CeCoIn$_5$ block layers and interact with the neighboring nonmagnetic or magnetic layers through the interface. 

 In CeCoIn$_5$/YbCoIn$_5$ superlattices, the superconductivity is strongly influenced by the local ISB at the interface, which seriously reduces the Pauli paramagnetic pair-breaking effect.  Our results demonstrate that the tricolor
Kondo superlattices provide a new playground for exploring exotic superconducting states in the strongly correlated 2D electron systems with the Rashba effect.

CeCoIn$_5$/CeRhIn$_5$  and CeCoIn$_5$/CeIn$_3$ superlattices, the superconducting and antiferromagnetic states coexist in spatially separated layers, but their mutual coupling via the interface significantly modifies the superconducting and magnetic properties.  In CeCoIn$_5$/CeRhIn$_5$ superlattices,  the superconductivity in the CeCoIn$_5$ BLs is profoundly affected by AFM fluctuations in the CeRhIn$_5$ BLs.  Upon suppressing the AFM order by applied pressure, the force binding superconducting electron pairs acquires an extreme strong coupling nature,  highlighting that the pairing interaction can be maximized by the critical fluctuations emanating from the magnetic QCP.   In CeCoIn$_5$/CeIn$_3$ superlattices, each CeIn$_3$ BL is magnetically coupled by RKKY  interaction through the adjacent CeCoIn$_5$ BLs.  In stark contrast to CeCoIn$_5$/CeRhIn$_5$ superlattices, the  superconductivity in the CeCoIn$_5$ BLs in CeCoIn$_5$/CeIn$_3$ superlattices is barely influenced by the AFM fluctuations in the CeIn$_3$ BLs, even when the CeIn$_3$ BLs are tuned in the vicinity of the AFM QCP by pressure.  The striking difference between the two Kondo superlattices provides direct evidence that 2D AFM fluctuations are essentially important for the pairing interactions in CeCoIn$_5$. 

Finally,  we describe the future prospect of the Kondo superlattices.  Cd- and Hg-doped CeCoIn$_5$ are known to show anomalous antiferromagnetic order at small doping levels\cite{Pham_2006, Nicklas_2007, Booth_2009}.
The fabrication of superlattices consisting of alternating layers of Cd- and Hg-doped CeCoIn$_5$ and pure CeCoIn$_5$ is expected to provide  important information on the interplay between superconductivity and unusual magnetic order.   The advantage of these superlattices is that the mismatch of the lattice constant is very small.    Recently, strongly correlated electron systems with strong Rashba interaction are attracting much attention\cite{Michishita_2019,Nogaki_2020,Wolf_2020}.  Thus fabricating superlattices including the noncentrosymmetric heavy fermion superconductors CePt$_3$Si\cite{Bauer_2004} and pressure-induced superconductors CeRhSi$_3$\cite{Kimura_2007}, CeIrSi$_3$\cite{Okuda_2007}, and heavy fermion superconductors whose inversion symmetry is broken locally around the Ce site CeRh$_2$As$_2$\cite{Khim_2021} will give important information of the interplay between superconductivity and strong Rashba interaction.  We also note that the  Dzyaloshinskii--Moriya interaction plays an important role for the Kondo effect and magnetic order at the interface between different materials. Theoretical research in this direction appears to be ongoing\cite{Shahzad_2017, Okumura_2018}, but the impact on the $f$-electron superlattice is not clear at the moment.

Besides the Kondo superlattices, the present MBE technique enables us to grow monolayer CeCoIn$_5$ and CeRhIn$_5$ on the top of the nonmagnetic YbCoIn$_5$ thin film.  It has been proposed theoretically that monolayer CeCoIn$_5$ forms a topological superconducting state owing to the strong Rashba interaction\cite{Yuan_2014,Daido_2017,Daido_2016,Read_2000}.  Most of the past research on topological superconductivity has focused on weakly correlated materials.  The topological superconductivity in monolayer CeCoIn$_5$ will give an opportunity to study topological excitations, including Majorana fermions in strongly correlated $d$-wave superconductors.    Moreover, it is an open question whether monolayer CeRhIn$_5$ exhibits magnetic order.   It is highly challenging to study these issues  by using the scanning tunneling microscopy  measurements.

The fabrication of a wide variety of Kondo superlattices paves a new way to study the entangled relationship between unconventional superconductivity and magnetism in strongly correlated electron systems, offering a route to exploring the emergence of novel superconducting systems and the roles of their interface.
 
\section*{Acknowledgments}
The authors acknowledge collaborations with Yongkang Luo, P. F. S. Rosa, F. Ronning, J. D. Thompson, K. Ishida, T. Yamanaka, G. Nakamine, S. Miyake, S. Nakamura, T. Ishii, R. Peters, Y. Tokiwa, Y. Kasahara, and T. Shibauchi. We thank H. Kontani and Y. Yanase for valuable discussion.
This work was supported by Grants-in-Aid for Scientific Research (KAKENHI) (No.\,25220710, No.\,18J10553, and No.\,18H05227) and Innovative Areas "Topological Material Science" (No.\,JP15H05852) and "3D Active-Site Science" (No.\,26105004) from the Japan Society for the Promotion of Science (JPSJ) and JST CREST (No.\,JP-MJCR18T2).

\section*{Data Availability}
The data that support the findings of this study are available from American Physical Society. Restrictions apply to the availability of these data, which were used under license for this study. Data are available from the authors upon reasonable request and with the permission of American Physical Society.

%

\end{document}